\documentclass[11pt,tightenlines,eqsecnum,floats,aps,amsmath,amssymb,prd,nofootinbib,shownopacs,floatfix]{revtex4-1}
\usepackage[english]{babel}
\usepackage[utf8x]{inputenc}
\usepackage{amsmath,amsfonts,amssymb,extarrows,mathrsfs}
\usepackage{xcolor}
\usepackage{physics}
\usepackage{graphicx}
\usepackage{float}
\usepackage{empheq}
\usepackage[colorlinks=false, pdfborder={0 0 0}]{hyperref}
\usepackage{bbm}
\usepackage{pdfpages}
\usepackage{makecell}
\usepackage{tensor}
\usepackage{dsfont}
\usepackage{mathrsfs}
\usepackage{mathtools}
\usepackage{varwidth}
\usepackage[toc,title,page]{appendix}
\usepackage[colorinlistoftodos,prependcaption,textsize=tiny]{todonotes}
\usepackage{titlesec}
\usepackage{tikz}
\usetikzlibrary{calc}
\usetikzlibrary{shapes.geometric, arrows}
\usetikzlibrary{shapes, positioning}
\usetikzlibrary{arrows.meta}
\usetikzlibrary{fit}
\usetikzlibrary{arrows.meta}
\usetikzlibrary{trees}
\usepackage{varwidth}

\tikzstyle{process} = [rectangle, rounded corners, minimum width=4.0cm, minimum height=1cm, draw=black, fill=blue!15]
\tikzstyle{arrow} = [thick, line width=0.6mm,->,>=latex]

\newcommand{\defeq}{\mathrel{\mathop:}=}
\newcommand{\eqdef}{\mathrel{\mathop=}:}
\renewcommand{\P}{\mathcal{P}}
\newcommand{\Q}{\mathcal{Q}}

\newcommand{\Lt}{\mathcal{L}(t)}
\renewcommand{\d}{\mathrm{d}}
\renewcommand{\r}{\hat{\rho}}
\newcommand{\rS}{\hat{\rho}^{(S^1)}_{\textrm{th}}}
\newcommand{\rI}{\hat{\rho}^{(I)}}

\newcommand{\Hil}{\mathcal{H}}
\newcommand{\HilS}{\mathcal{H}_S}
\newcommand{\HilE}{\mathcal{H}_\varepsilon}
\newcommand{\Htot}{\hat{H}_{\mathrm{tot}}}
\newcommand{\Hint}{\hat{H}_{\mathrm{int}}}
\newcommand{\Hn}{\hat{H}_{0}}
\renewcommand{\thesection}{\Roman{section}}
\renewcommand{\thesubsection}{\Alph{subsection}}

\makeatletter
\renewcommand{\p@subsection}{\thesection.} 
\renewcommand{\p@subsubsection}{\thesection.\thesubsection.} 
\AtBeginDocument{\let\LS@rot\@undefined} %

\begin{document}
\title{An open scattering model in polymerized quantum mechanics}
\author{Kristina Giesel}
\thanks{kristina.giesel@gravity.fau.de}
\author{Michael Kobler}
\thanks{michael.kobler@gravity.fau.de}
\affiliation{Institute for Quantum Gravity,Theoretical Physics III, Department of Physics,  \\ Friedrich-Alexander-Universität Erlangen-N\"urnberg,
Staudtstr. 7, 91058 Erlangen, Germany}

\begin{abstract}
We derive a quantum master equation in the context of a polymerized open quantum mechanical system for the scattering of a Brownian particle in an ideal gas environment. The~model is formulated in a top-down approach by choosing a Hamiltonian with a coupling between the system and environment that is generally associated with spatial decoherence. We extend the existing work on such models by using a non-standard representation of the canonical commutation relations, inspired by the quantization procedure applied in loop quantum gravity, which yields a model in which position operators are replaced by holonomies. The~derivation of the master equation in a top-down approach opens up the possibility to investigate in detail whether the assumptions, usually used in such models in order to obtain a tractable form of the dissipator, hold also in the polymerized case or whether they need to be dropped or modified. Furthermore, we discuss some physical properties of the master equation associated to effective equations for the expectation values of the fundamental operators and compare our results to the already existing models of collisional~decoherence.
\end{abstract}

\maketitle

\newpage
\section{Introduction} \label{sec:introduction}
Decoherence and the quantum-to-classical transition are often discussed in terms of open quantum systems and in consequence formulated as a master equation, often in the form of the celebrated Lindblad equation \cite{Lindblad:Semigroups,Gorini,Breuer:OpenSystems,Breuer:Concepts, joos2013decoherence, zurek2006decoherence}. In this framework, the system of interest is usually coupled to an environment with a considerably higher number of degrees of freedom, expressed in terms of an interaction Hamiltonian. Quantum mechanical models require a \textit{choice} of interaction Hamiltonian \cite{Caldeira:QBM,Schlosshauer:Boson-Spin}, whereas in the case of field theories, the interaction Hamiltonian is often provided directly by the canonical formalism that is required for quantization, see for instance \cite{Breuer:DecoherenceQED,oniga2016quantum, anastopoulos2013master,Fahn:2022zql}. The~second-order (Born-Markov) master equation is consequently derived by averaging over the environmental degrees of freedom, which only enter the master equation in an effective manner via autocorrelation functions. Next to the Lindblad equation, this is the most frequently used approach of a time-local equation, which however is not necessarily trace-preserving nor automatically completely positive for arbitrarily large time scales \cite{homa2019positivity}. The~Lindblad equation requires the generator of the associated dynamical semigroup to be bounded \cite{Lindblad:Semigroups,Gorini}, however, which is rarely the case in physical applications. Whether a model exhibits a Lindblad-like master equation depends on a multitude of assumptions as we will discuss in the course of this work. Decoherence has been studied for a large variety of quantum mechanical and quantum field theoretical models, from collisional decoherence of a Brownian particle in an ideal quantum gas environment \cite{joos1985emergence,hornberger2003collisional,busse2009pointer,hornberger2006master, hornberger2008monitoring} to Quantum Brownian Motion (QBM) of the quantum harmonic oscillator \cite{Caldeira:QBM} in a bath of bosonic modes over the spin-boson model \cite{Breuer:OpenSystems,clos2012quantification,Schlosshauer:DecoherenceBook} to an oscillator system coupled to a collection of two-level systems \cite{Schlosshauer:Boson-Spin}. Field theoretical applications for example include decoherence through Bremsstrahlung in QED \cite{Breuer:DecoherenceQED}, decoherence of cosmological perturbations \cite{hollowood2017decoherence,boyanovsky2015effective,laflamme1991reduced,polarski1996semiclassicality,halliwell1989decoherence} and decoherence of matter degrees of freedom due to the interaction with the gravitational field in both ADM and Ashtekar variables \cite{anastopoulos2013master,oniga2016quantum,Fahn:2022zql}.
Loop Quantum Gravity (LQG) as well as Loop Quantum Cosmology (LQC) are known to use a so-called polymer representation for their elementary variables \cite{Thiemann:MCQGR,rovelli2004quantum,Agullo:LQCReview,Ashtekar:LQCReport, Bojowald:LQCStructure}. This~choice is strongly related to the underlying symmetries in LQC and LQG respectively. The~quantum mechanical version of this representation turns out to be unitarily inequivalent to the ordinary Schrödinger representation. The~vast majority of decoherence models deals with the usual Fock quantization methods, and the effect of polymer quantization on open quantum systems is yet to be fully explored. In the context of gravitational decoherence, this seems to be of paramount importance, since in the framework of General Relativity (GR) everything couples to gravity and hence it can be seen as a reasonable choice for an environment of the remaining matter degrees of freedom.  There has been previous work on polymerized thermodynamics \cite{chacon2011statistical} and open quantum (mechanical) systems \cite{feller2017surface}. The~vast majority of canonical models in the existing literature, such as collisional decoherence \cite{joos1985emergence,hornberger2003collisional,hornberger2006master}, Quantum Brownian Motion (QBM) \cite{Caldeira:QBM} and spin-boson applications \cite{Schlosshauer:Boson-Spin,Breuer:OpenSystems,Schlosshauer:DecoherenceBook} have not been formulated and investigated in the context of polymerized quantum mechanics to the knowledge of the authors.
\\

The~derivation and analysis of master equations associated to (constrained) quantum field theories already is a complicated task in the context of a Fock quantization scheme, see for instance  \cite{anastopoulos2013master,oniga2016quantum,Fahn:2022zql} for the formulation of concrete models. The~complexity further increases if, instead of the Fock representation, a non-standard representation such as the polymer representation as in LQG is involved. In order to obtain a better knowledge of how a given choice of representation affects a given decoherence model, we consider such models in the context of polymerized quantum mechanics as a first step towards formulating decoherence models inspired by LQC and LQG, respectively. In the case of LQC, another interesting aspect regarding the quantum-to-classical transition arises. Most already existing models in LQC are based upon the framework of closed quantum systems. The~classical singularity, present in for instance FLRW models, is resolved and replaced by a bounce. It has been shown that if one starts with a semiclassical state representing a classical FLRW universe today and evolves backwards, the evolution follows the classical trajectory until a critical matter density is reached and then bounces, joining in the pre-bounce era a trajectory which was classically headed towards the big crunch. This is in stark contrast to the Wheeler--DeWitt theory that is based on a Schrödinger representation. An interesting question is whether the decoherence formalism could provide additional justification for the semiclassical quantum states used in LQC, see for instance the work in \cite{Ashtekar:2007robustness}, hence further supporting the robustness of the key features of LQC from an open quantum systems perspective.  
\\

In this work we derive a collisional decoherence model in the framework of polymerized quantum mechanics, that we denote as an open polymerized scattering model, the latter notion based on the structure of the chosen total Hamiltonian and analyse its physical properties. The~main differences to the approaches in previous works on collisional decoherence, that is the random scattering of a Brownian particle in a thermal environment, see for instance \cite{joos1985emergence,busse2009pointer,hornberger2003collisional,hornberger2006master}, is on the one hand that we work on the polymerized Hilbert space. For our choice of representation, the position operator does not exist but the corresponding holonomies do and thus its analogue is expressed in terms of a difference of holonomy operators. Additionally, we start with a given total Hamiltonian and derive the corresponding master equation from it, whereas the derivation in \cite{hornberger2003collisional,hornberger2006master,hornberger2008monitoring} holds for large system masses $M \gg m$ as compared to the environment and follows an S-matrix approach in the derivation of the master equation. In this context, the chosen basis for the thermal density operator (the initial state) of the environment is crucial in the avoidance of singularities within the environmental autocorrelation functions \cite{hornberger2003collisional}. As we will show, these singularities are circumvented in the model considered in this article by virtue of a polymerization of the position operator. A comparison with the solutions obtained from collisional decoherence using Schrödinger quantization shows that the solutions obtained here differ in characteristic features. This is in a sense expected because the corresponding Schrödinger-like model, due to its singularities, cannot be derived in the same way. The~resulting effective differential equations for the expectation values of momentum and (polymerized) position do not exhibit any divergencies either and can be solved analytically in lowest order of the polymerization parameter.  We further discuss the applicability of the second Markov approximation, often implemented alongside a rotating-wave approximation \cite{fleming2010rotating} in order to retrieve a master equation of Lindblad-type. Interestingly, the Lindblad operators in the final master equation in \cite{hornberger2003collisional,hornberger2006master,hornberger2008monitoring} looked similar to $U(1)$ holonomies found for example in the discussion of coherent states \cite{Bahr:Coherent} or the quantization of the canonical pair of angle and angular momentum \cite{Kastrup:AnglePair,Kastrup:Optical}, respectively. The~Hilbert spaces in these latter works exactly correspond to a superselection sector of the polymerized Hilbert space encountered in LQC \cite{corichi2007polymer}. However, to the understanding of the authors, it seems to be not straightforward to find a form of an underlying total Hamiltonian that yields the same final Lindblad equations obtained in \cite{hornberger2003collisional,hornberger2006master,hornberger2008monitoring} in the course of the derivation of a Born--Markov master equation.
\\

The~paper is structured as follows: For the benefit of the reader and to make the article self-contained, we briefly review the Nakajima--Zwanzig formalism \cite{Zwanzig:Ensemble,Nakajima:Diffusion} in Section \ref{sec:master_review} and the associated projection-operator technique in Section \ref{sec:ReviewPOT}. In Section \ref{sec:Review:BMA}, largely following~\cite{Breuer:Concepts, Breuer:Timelocal}, the ensuing, and in principle exact, now time-local master equation is then further simplified by a number of assumptions and/or approximations, most prominently the Born- and Markov approximations, respectively. We are particularly interested in understanding which specific assumptions enter into the individual steps of the derivation and see whether those can be carried over to derivation of the polymerized model. In Section \ref{sec:DerPolyModel} we derive the polymerized open scattering model using a top-down approach, that requires an initial choice of a total Hamiltonian encoding the dynamics of system and environment as well as their interaction. Again for the benefit of the reader and in order to introduce our notation, we provide a concise discussion of the different polymer representations, their properties and the role of superselection sectors in Section \ref{sec:U1model}. The~individual steps of the detailed derivation of the model are presented in Section \ref{sec:model_application}. An investigation and discussion about the physical properties of the model can be found in Section \ref{sec:model_interpretation} where we point out the similarities and key differences in regard to the established models of collisional decoherence. This is done in terms of the differential equations for the expectation values of the canonical variables and their associated solutions. These solutions are obtained after we truncate at a certain order in the polymerization parameter and thus decouple the differential equations at lowest order. Finally, in Section \ref{sec:conclusion} we summarise our results, relate them to the already existing different approaches for collisional decoherence and discuss possible generalisation of the open polymerized scattering model in various~directions.
\section{Review of the Derivation of Quantum Master Equations} \label{sec:master_review}

In the theory of open quantum systems one usually resorts to a statistical treatment of the many degrees of freedom involved, often in terms of density operators. The~main difference to closed quantum systems is that one considers a coupling of the system involving the relevant degrees of freedom under consideration to a chosen environment. The~goal in the context of quantum master equations is to derive an effective first-order equation of motion for the relevant  degrees of freedom without requiring prior knowledge of the full dynamics of the environment, see for instance the textbooks \cite{Breuer:OpenSystems, joos2013decoherence}. Depending on the level of complexity of the dynamics of the total system and different approximative techniques, the resulting so-called master equations exhibit a variety of properties, the most well-known being of the Lindblad type \cite{Lindblad:Semigroups,Gorini} which has the following (off-diagonal) form:

\begin{align}
    \frac{\d}{\d t} \hat{\rho}_S(t) = -i [\hat{H}_S, \hat{\rho}_S(t)] + \frac{1}{2} \sum_{\alpha, \beta \in \mathcal{I}} \Gamma_{\alpha \beta} \Big( \big[ \hat{S}_\alpha, \hat{\rho}_S(t) \hat{S}^\dagger_\beta \big] + \big[ \hat{S}_\alpha \hat{\rho}_S(t), \hat{S}^\dagger_\beta \big] \Big),
\end{align}
where $\hat{H}_S$ is the Hamiltonian of the system which represents the unitary part of the time evolution, $\mathcal{I}$ is a model-dependent, most commonly finite index set, the $\Gamma_{\alpha \beta}$ are time-indepent, real coefficients and $\hat{S}_\alpha, \hat{S}^\dagger_\beta$ are time-independent operators on the Hilbert space of the system of interest. Due to the positive semi-definiteness of $\Gamma$ we can bring the Lindblad equation into a diagonal form via a unitary transformation, the $\Lambda_\alpha$ are the corresponding eigenvalues of $\Gamma$.

\begin{align}
    \frac{\d}{\d t} \hat{\rho}_S(t) = -i [\hat{H}_S, \hat{\rho}_S(t)] + \frac{1}{2} \sum_{\alpha \in \mathcal{I}} \Lambda_{\alpha} \Big( \big[ \hat{L}_\alpha, \hat{\rho}_S(t) \hat{L}^\dagger_\alpha \big] + \big[ \hat{L}_\alpha \hat{\rho}_S(t), \hat{L}^\dagger_\alpha \big] \Big),
\end{align}
with Lindblad operators $\hat{L}_\alpha = \sum_\beta h_{\alpha \beta} \hat{S}_\beta$ obtained via a change of basis $h$ associated to the diagonalization of $\Gamma$. This equation guarantees a completely positive and trace-preserving evolution of the system's density operator, while in a more general setting, this is no longer true. In cases where the interaction with the environment is not Markovian in nature or if the initial condition is non-separable (that is, a non-vanishing initial entanglement entropy), it might not be possible to reach a Lindblad form at all \cite{Breuer:Generalization,Megier:Interplay}. This does however not mean that complete positivity or trace preservation is violated \cite{Schlosshauer:DecoherenceBook}, these properties rather need to be checked once the master equation for a given system of interest is derived. We will make a clear distinction between and consecutively comment on the various types of master equations from the more general perspective of projection superoperators in the next sections.

\subsection{Brief Review of the Projection Operator Techniques}
\label{sec:ReviewPOT}
Exact projection superoperator techniques in conjunction with integral master equations in the context of open quantum systems were first applied by Nakajima and Zwanzig in \cite{Nakajima:Diffusion} and \cite{Zwanzig:Ensemble}, respectively. In this chapter however, we will largely follow the more pedagogical outline given in \cite{Breuer:OpenSystems}, where we fill in some additional steps that we deem important for the uninitiated reader. The~starting point is a system that can be split into a system and environmental part that are labelled by $S$ and $\varepsilon$ respectively.  The~corresponding total Hamiltonian for  a given model $\Htot = \Hn + \alpha \Hint$ acts on the total Hilbert space $\Hil = \HilS \otimes \HilE$, where $\Hn = \hat{H}_S \otimes \mathds{1}_\varepsilon + \mathds{1}_S \otimes \hat{H}_\varepsilon$ describes the free evolution of the system and the environment, and $\Hint$ accounts for the interaction between both.  Here $\alpha$ is a (dimensionless) coupling constant without, at this point, any further assumptions regarding its magnitude. Note that the coupling will not be dimensionless in the case of the model considered here, however within the scope of this section it will make the derivation more transparent, since we do not need to split the interaction Hamiltonian into the system and environment, respectively. This split is performed at the level of the Born--Markov master equation. The~Banach space of density operators of the total system is given by the linear, positive, self-adjoint and unit trace operators with trace norm on the combined Hilbert space: 

\begin{align}
    \mathcal{D}(\mathcal{H}) \defeq \big\{ \r: \Hil \to \Hil \; \textrm{linear}: \; \norm{\r}_1 \leq 1, \Tr(\r) = 1, \r \geq 0, \r^\dagger = \r \big\},
\end{align}
where $\norm{\r}_1 \defeq \sqrt{\Tr(\r^\dagger \r)}$ is the trace norm. The~time-evolution of the total density operator in the interaction picture is encoded in the Liouville--von Neumann equation:

\begin{equation}
    \frac{\d}{\d t} \rI(t) = -i \alpha [\Hint(t), \rI(t)] = \alpha \Lt \rI(t),
    \label{eq:Liouville}
\end{equation}
with the Liouville superoperator $\mathcal{L}(t)$ and the transformation given by the free evolution:

\begin{equation}
\rI = \hat{U}_0^\dagger(t,t_0) \, \r(t_0) \, \hat{U}_0(t,t_0), \quad \hat{U}_0(t,t_0) = \mathcal{T}_{\leftarrow} \exp \Big\{-i \int\limits_{t_0}^t \d \tau \, \Hn(\tau) \Big \},
\end{equation}
with $\mathcal{T}_{\leftarrow}$ depicting the time-ordering meta-operator, the largest argument is ordered to the left. In order to be able to obtain a closed differential equation for the evolution of the relevant degrees of freedom, we introduce projection superoperators acting on the space of density operators of the total system

\begin{align}
    \P&: \mathcal{D}(\mathcal{H}) \to \mathcal{D}(\mathcal{H}), \; \r \mapsto \P \r = \Tr_\varepsilon (\r) \otimes \r_\varepsilon = \r_S \otimes \r_\varepsilon \label{eq:P_projector} \\
    \Q&: \mathcal{D}(\mathcal{H}) \to \mathcal{D}(\mathcal{H}), \; \r \mapsto \Q \r = \r - \P \r = \r - \r_S \otimes \r_\varepsilon\, . \label{eq:Q_projector}
\end{align}

Here $\P,\Q$ are constructed in a way that $\P$ projects on the relevant part and $\Q$ on the irrelevant part of the total system, $\r_S$ and  $\r_\varepsilon$ depict the relevant and irrelevant part, respectively.  The~properties $\P + \Q = \mathds{1}_{\mathcal{D}(\mathcal{H})}$, $\P^2=\P$, $\Q^2=\Q$ and $\P \Q = \Q \P = 0$ are elementary given the definitions above. We would like to point out that $\r_S$ is not solely the system's density operator but already contains information about the correlation between the system and its environment. This conceptual detail in notation is especially relevant in the context of non-separable initial conditions as we will see in the course of this section. Application of the individual operators to the Liouville--von Neumann Equation \eqref{eq:Liouville} and taking into account the split of the identity operator in terms of $\Q$ and $\P$ yields:

\begin{align}
    \P \frac{\d}{\d t} \rI(t) &= \frac{\d}{\d t} \P \rI(t) = \alpha \P \Lt \rI(t) = \alpha \big( \P \Lt \P \rI(t) + \P \Lt \Q \rI(t) \big), \label{eq:RelPart}\\
    \Q \frac{\d}{\d t} \rI(t) &= \frac{\d}{\d t} \Q \rI(t) = \alpha \Q \Lt \rI(t) = \alpha \big( \Q \Lt \P \rI(t) + \Q \Lt \Q \rI(t) \big)\, . \label{eq:IrrPart}
\end{align}

Note the relative positions of the projectors with respect to $\Lt$ and $\rI(t)$. These are \textit{not} equations in the variables $\P \rI(t)$ and $\Q \rI(t)$, respectively, and hence need to be treated slightly differently than the standard case of a time-ordered, exponentiated Liouvillian. The~strategy to obtain a closed form of the equation for $\P \rI(t)$ is to formally solve for $\Q \rI(t)$ and reinsert the obtained expression into the equation for the relevant part, after which appropriate approximations can be implemented. The~solution for the irrelevant part is given by:

\begin{align}
\label{eq:SolIrrPart}
    \Q \rI(t) &= \mathcal{G}(t,t_0) \Q \rI(t_0)
    + \alpha \int\limits_{t_0}^t \d s \, \mathcal{G}(t,s) \Q \mathcal{L}(s) \P \rI(s),
\end{align}
where the propagator $\mathcal{G}(t,t_0)$ is given by 

\begin{align*}
    \mathcal{G}(t,t_0) &= \mathcal{T}_{\leftarrow} 
    \exp \Big\{ \alpha \int\limits_{t_0}^t \d \tau \, \Q \mathcal{L}(\tau) \Big \},\quad 
    \mathcal{G}(t_0,t_0)= \mathds{1}_{\mathcal{D}(\mathcal{H})}
\end{align*}
and satisfies the differential equation
\begin{align*}
\frac{\d}{\d t} \mathcal{G}(t,t_0) = \alpha \Q \Lt \mathcal{G}(t,t_0).    
\end{align*}

An explicit time derivative and the fact that $\mathcal{G}(t,t) = \mathds{1}_{\mathcal{D}(\mathcal{H})}$ proves that this is the unique solution. As a last step towards an exact master equation we insert the formal solution of $\Q \rI(t)$ in \eqref{eq:SolIrrPart} into the equation for the relevant part \eqref{eq:RelPart} of the total \mbox{density~operator:}

\begin{align}
    \frac{\d}{\d t} \P \rI(t) &= \alpha \P \Lt \P \rI(t) + \alpha \P \Lt \mathcal{G}(t,t_0) \Q \rI(t_0) \nonumber \\
    &+ \alpha^2 \int\limits_{t_0}^t \d s \, \P \Lt \mathcal{G}(t,s) \Q \mathcal{L}(s) \P \rI(s)
    \label{eq:NZ}
\end{align}

This is the celebrated \textit{Nakajima-Zwanzig} equation, it is exact and was obtained without the assumption of any approximations. However, the solution is equally difficult to obtain as the dynamics for the total system. This is mostly due to the fact that the convolution in the second line  in  \eqref{eq:NZ} is non-local in the temporal coordinate because of non-Markovian memory effects.

If we further assume that the interaction Hamiltonian is of the form $\sum_\alpha \hat{S}_\alpha \otimes \hat{E}_\alpha$ for system's and environmental operators $\hat{S}_\alpha, \hat{E}_\alpha$ respectively, the latter composed of linear combinations of creation- and annihilation-like variables only, as is the case for the canonical models \cite{Caldeira:QBM,Schlosshauer:Boson-Spin}, we can further simplify Equation \eqref{eq:NZ} by a closer examination of the firs term. Insertion of the definition of the projector $\P$ on contributions with an odd number of Liouvillians yields:

\begin{align}
    \P \mathcal{L}(t_1) \ldots &\mathcal{L}(t_{2n+1}) \P \rI(t) = 
    \P \mathcal{L}(t_1) \ldots \mathcal{L}(t_{2n+1}) \big( \Tr{\rI(t)} \otimes \hat{\rho}_\varepsilon \big) \label{eq:odd_liouvillian}\\
    &= (-i)^{2n+1} \P \big[\Hint(t_1), \big[ \Hint(t_2), \ldots, \big[\Hint(t_{2n+1}), \big( \Tr{\rI(t)} \otimes \hat{\rho}_\varepsilon \big) \big] \ldots \big]\nonumber \\
    &=  (-i)^{2n+1} \Tr{ \big[\Hint(t_1), \ldots, \big[\Hint(t_{2n+1}), \big( \Tr{\rI(t)} \otimes \hat{\rho}_\varepsilon \big) \big] \ldots \big]} \otimes \hat{\rho}_\varepsilon.\nonumber
\end{align}

It is obvious that an uneven number of interaction Hamiltonians of the assumed form leads to a vanishing contribution. Hence, the Nakajima--Zwanzig equation with a non-Markovian memory kernel $\mathcal{K}(t,s)$ for a broad class of models can be rewritten as:

\begin{align*}
     \frac{\d}{\d t} \P \rI(t) = 
     &\alpha \P \Lt \mathcal{G}(t,t_0) \Q \rI(t_0)
    + \int\limits_{t_0}^t \d s \, \mathcal{K}(t,s) \P \rI(s), \\
    &\mathcal{K}(t,s) \defeq \alpha^2 \P \Lt \mathcal{G}(t,s) \Q \mathcal{L}(s).
\end{align*}

While in principle exact, the Nakajima--Zwanzig equation is not a feasible way to obtain the effective dynamics in a reasonable fashion due to its complicated convolution structure. This is why we would rather prefer a time-local generator $K(t)$, which indeed can be obtained if we are willing to take more stringent assumptions regarding the model considered here, all of which will be mentioned at the time they need to be imposed. \mbox{To this} end, we consider the two propagators $F(t,s)$ and $G(t,s)$ such that

\begin{align*}
G(t,s) \rI(s) = \rI(t)\quad{\rm and}\quad  F(t,s) \circ G(t,s) = \mathds{1}_{\mathcal{D}(\mathcal{H})}, 
\end{align*}
then we have

\begin{align}
    \frac{\d}{\d t} F(t,s) \circ G(t,s) \rI(s) = \Big( \frac{\d}{\d t} F(t,s) \Big) \rI(t) + \alpha F(t,s) \Lt \rI(t) = 0,
\end{align}
which can be directly solved for the inverse propagator $\frac{\d}{\d t} F(t,s) = - \alpha F(t,s)$ leading~to:

\begin{align}
    F(t,s) = \mathcal{T}_\rightarrow \exp \Big\{ - \alpha \int\limits_s^t \d \tau \, \mathcal{L}(\tau) \Big\}.
\end{align}

Here we used $\mathcal{T}_\rightarrow$ for anti-time-ordering. Note that $\mathcal{G}(t,s) \neq G(t,s)$, as the latter is the total propagator of the combined system, i.e., the time-ordered exponential of the integrated Liouvillian $\Lt$. Now $\rI(s) = F(t,s) (\P + \Q) \rI(t)$, which can be inserted into the solution for the irrelevant part in \eqref{eq:SolIrrPart}:

\begin{align}
    \Q \rI(t) &= \mathcal{G}(t,t_0) \Q \rI(t_0) + \alpha \int\limits_{t_0}^t \d s \, \mathcal{G}(t,s) \Q \mathcal{L}(s) \P F(t,s) (\P + \Q) \rI(t) \\
    &\eqdef \mathcal{G}(t,t_0) \Q \rI(t_0) + \Sigma(t,t_0) ( \P + \Q) \rI(t).\nonumber
\end{align}
with the corresponding definition for $\Sigma(t,t_0)$ that can be read of from the second line. A~simple rearrangement of terms reveals that

\begin{align}
    \big( \mathds{1}_{\mathcal{D}(\mathcal{H})}
    - \Sigma(t,t_0) \big) \Q \rI(t) = \mathcal{G}(t,t_0) \Q \rI(t_0) + \Sigma(t,t_0) \P \rI(t). \label{eq:one_minus_sigma}
\end{align}

The~strategy to obtain an explicit expression for $\Q \rI(t)$ from the equation above is to solve for $\Q \rI(t)$ by subsequently inverting the superoperator $\mathds{1}_{\mathcal{D}(\mathcal{H})} - \Sigma(t,t_0)$, after which we proceed to insert the obtained expression into our formal solution for the relevant projection in \eqref{eq:SolIrrPart}. We know by definition $\Sigma(t_0,t_0) = 0$ and $\Sigma(t,t_0)|_{\alpha = 0} = 0$, hence one can show \cite{Fahn:2022zql} that $\mathds{1}_{\mathcal{D}(\mathcal{H})} - \Sigma(t,t_0)$ is invertible for sufficiently small time scales $t - t_0$ and appropriate, that is weak to intermediate coupling constant $\alpha$. The~larger the coupling, the smaller is in general the time-scale for which $\Sigma(t,t_0)$ can be inverted. The~\textit{time-local} and \textit{convolutionless} (TCL) generator obtained in this way can then be expanded in a perturbative series in terms of powers of the coupling constant. We obtain

\begin{align*}
    \Q \rI(t) = \big( \mathds{1}_{\mathcal{D}(\mathcal{H})} - \Sigma(t,t_0) \big)^{-1} \mathcal{G}(t,t_0) \Q \rI(t_0) 
    + \big( \mathds{1}_{\mathcal{D}(\mathcal{H})} - \Sigma(t,t_0) \big)^{-1} \Sigma(t,t_0) \P \rI(t).
\end{align*}

The~ansatz for the inverse superoperator is given by a geometric expansion, which reproduces the properties mentioned above and allows for the possibility to rediscover the standard form of the master equation one obtains by simply tracing out the irrelevant degrees of freedom in the Liouville--von Neumann equation. Concretely, one uses the following expansion:

\begin{align}
\big( \mathds{1}_{\mathcal{D}(\mathcal{H})} - \Sigma(t,t_0) \big)^{-1} = \sum_{n=0}^\infty \big( \Sigma(t,t_0) \big)^n, \quad \Sigma(t,t_0) = \sum_{m=1}^\infty \alpha^m \Sigma_m(t,t_0).
\end{align}

The~fact that the second sum starts at $m=1$ is simply because at lowest order in $\alpha$, the expression for  $\Sigma(t,t_0)$ is linear. The~contributions $\Sigma_m(t,t_0)$ label the individual orders in this ansatz. The~particular choice of notation will be readdressed once we give an explicit example. This result combined with the Nakajima--Zwanzig equation \eqref{eq:NZ} yields the final equation for the relevant part also known as the time-convolutionless (TCL) master equation in the literature:

\begin{align} 
    \frac{\d}{\d t} \P \rI(t) &= \alpha \P \Lt \P \rI(t) + \alpha \P \Lt \big( \mathds{1}_{\mathcal{D}(\mathcal{H})} - \Sigma(t,t_0) \big)^{-1} \Big( \mathcal{G}(t,t_0) \Q \rI(t_0) + \Sigma(t,t_0) \P \rI(t) \Big) \nonumber \\[0.35em]
    &= \big( \alpha \P \Lt \P + \alpha \P \Lt \big( \mathds{1}_{\mathcal{D}(\mathcal{H})} - \Sigma(t,t_0) \big)^{-1} \Sigma(t,t_0) \big) \P \rI(t) + \mathcal{I}(t,t_0) \Q \rI(t_0) \nonumber \\[0.25em]
    &= \big( \alpha \P \Lt \P + \alpha \P \Lt \sum_{n=0}^\infty \big( \Sigma(t,t_0) \big)^{n+1} \big) \P \rI(t) + \mathcal{I}(t,t_0) \Q \rI(t_0) \nonumber \\[0.15em]
    &= \alpha \P \Lt \sum_{n=0}^\infty \big( \Sigma(t,t_0) \big)^n \P \rI(t) + \mathcal{I}(t,t_0) \Q \rI(t_0) \nonumber \\[0.25em]
    &= \alpha \P \Lt \sum_{n=0}^\infty \Big( \sum_{m=1}^\infty \alpha^m \Sigma_m(t,t_0) \Big)^n \P \rI(t) + \mathcal{I}(t,t_0) \Q \rI(t_0) \nonumber \\[0.25em]
    &\eqdef K(t,t_0) \P \rI(t) + \mathcal{I}(t,t_0) \Q \rI(t_0),
    \label{eq:TCL}
\end{align}

where the definition of $\mathcal{I}(t,t_0)$ can be read of from the second equality. Given the assumptions in order to perform all the necessary steps, this form of the master equation is much more tractable than the Nakajima--Zwanzig equation itself with the involved convolution kernel. Note, that furthermore, albeit the time-local nature of Equation \eqref{eq:TCL}, the underlying processes can be non-Markovian in nature. The~TCL master equation admits an expansion of the homogeneous and inhomogeneous contributions in terms of \textit{ordered cumulants} due to a result of van Kampen \cite{vanKampen:I,vanKampen:II}. This perturbative expansion is a valid possibility to tackle non-Markovian dynamics of open quantum systems in a time-local form at any order of the coupling parameter and is hence used in field theory applications of decoherence theory \cite{Breuer:DecoherenceQED} alongside the phase influence functional approach \cite{Breuer:Timelocal,Breuer:Concepts}. For the purpose of this work it is sufficient to consider the lowest non-vanishing order of this expansion which corresponds to an idealized Markov process. This will be made more clear in the next subsection alongside with the Born approximation of an exactly separable system.

\subsection{Brief Review of the Born Approximation and the Markov Processes in the Context of \mbox{Master Equations}}
\label{sec:Review:BMA}
In order to arrive at a tractable master equation for the total system in Section \ref{sec:U1model}, we impose separability for all times and a perfect Markov process. By virtue of the definition of the irrelevant projector $\Q$, see Equation (\ref{eq:Q_projector}), it is elementary that the inhomogeneous part vanishes if the initial condition is separable, that is $\rI(0) = \rI_S(0) \otimes \rI_\varepsilon(0)$ and consequently $\P \rI(0) = \rI(0)$. If we furthermore assume that the environment is in a thermal state, that is the environmental density operator commutes with its associated Hamiltonian, $[\hat{H}_\varepsilon, \r_\varepsilon(t)] = 0$ for all times it is clear that within a reasonable approximation, we can write the total density operator as $\rI(t) = \rI_S(t) \otimes \rI_\varepsilon (0)$. In this case, the second-order TCL master equation can be rewritten in terms of integrals involving environmental correlation functions alongside with the interaction picture's system operators. According to \eqref{eq:TCL} the form of the TCL master equation up to second order reads:

\begin{align}
    \frac{\d}{\d t} \P \rI(t) &= \alpha^2 \int\limits_{t_0}^t d\tau \, \P \Lt \Q \mathcal{L}(\tau) \P \rI(t) + \mathcal{O}(\alpha^4) \\
    &= - \alpha^2 \int\limits_{t_0}^t d\tau \,  \Tr_\varepsilon \Big( \big[ \Hint(t), \big[ \Hint(\tau),  \Tr_\varepsilon \big( \rI(t) \big) \otimes \hat{\rho}_\varepsilon \big] \big] \Big) \otimes \hat{\rho}_\varepsilon + \mathcal{O}(\alpha^4) \nonumber\\
    &= - \alpha^2 \int\limits_{t_0}^t d\tau \,  \Tr_\varepsilon \Big( \big[ \Hint(t), \big[ \Hint(\tau),  \rI_S(t) \otimes \hat{\rho}_\varepsilon \big] \big] \Big) \otimes \hat{\rho}_\varepsilon + \mathcal{O}(\alpha^4). \nonumber
\end{align}

In the first line we used that $\mathcal{Q} = \mathds{1} - \mathcal{P}$ and the fact that contributions with an odd number of Liouvillians (\ref{eq:odd_liouvillian}) vanish based on the assumed form of our interaction Hamiltonian. The~second step was to insert the definition of the relevant projector (\ref{eq:P_projector}), after which we used that per definition $\Tr_\varepsilon \big( \rI(t) \big) = \rI_S(t)$. An important detail to notice here is that the inner commutator bracket is evaluated with $\rI_S(t) \otimes \hat{\rho}_\varepsilon$ rather than $\rI_S(\tau) \otimes \hat{\rho}_\varepsilon$ as it would be the case if we directly evaluate the second-order expansion of the full Nakajima--Zwanzig equation from \eqref{eq:NZ}. Both equations are of second order and are expected to describe the dynamics of the total system with a comparable accuracy \cite{Breuer:OpenSystems}. However one of them is time-local while the other one is not. In the framework of the Markov approximation, that is a memory-less environment, the transition from the time convolution to the second-order TCL form is often done by hand \cite{Brasil:Simple}, showing that the difference $\rI(t) - \rI(\tau)$ is of at least second order in $\alpha$. In other words, non-Markovian corrections occur in the order $\alpha^4$ and higher, where the full expansion to infinite order yields the exact dynamics for both the non-local Nakajima--Zwanzig equation and the TCL master equation, respectively. The~exactness of the latter is only limited by the invertibility of $\big(\mathds{1} - \Sigma(t,t_0)\big)$ given by Equation (\ref{eq:one_minus_sigma}) from the previous section, which in turn depends on the coupling strength. Intuitively speaking, the truncation after second order in $\alpha$ is justified whenever the correlation times of the environment are much shorter than the ones of the system, e.g., in the case of a high-temperature thermal bath \cite{Schlosshauer:DecoherenceBook}. Given that the interaction Hamiltonian has the previously assumed form $\Hint(t) = \sum_\alpha \hat{S}_\alpha(t) \otimes \hat{E}_\alpha(t)$ we can reformulate the second-order TCL master equation as, where we choose to neglect terms of order $\mathcal{O}(\alpha^4)$ from now on:

\begin{align}
    \frac{\d}{\d t} \P \rI(t) &\approx - \alpha^2 \int\limits_{t_0}^t d\tau \,  \sum_{\alpha , \beta \in \mathcal{I}} \Tr_\varepsilon \Big( \big[ \hat{S}_\alpha(t) \otimes \hat{E}_\alpha(t), \big[ \hat{S}_\beta(\tau) \otimes \hat{E}_\beta(\tau),  \rI_S(t) \otimes \hat{\rho}_\varepsilon \big] \big] \Big) \otimes \hat{\rho}_\varepsilon \\
    &= - \alpha^2 \int\limits_{t_0}^t d\tau \,  \sum_{\alpha , \beta \in \mathcal{I}} \Big[ \hat{S}_\alpha(t) \hat{S}_\beta(\tau) \rI_S(t) \Tr_\varepsilon \Big( \hat{E}_\alpha(t) \hat{E}_\beta(\tau) \hat{\rho}_\varepsilon \Big) \nonumber\\
    &\qquad \qquad \qquad \qquad - \hat{S}_\alpha(t) \rI_S(t) \hat{S}_\beta(\tau) \Tr_\varepsilon \Big( \hat{E}_\alpha(t) \hat{\rho}_\varepsilon \hat{E}_\beta(\tau) \Big) \nonumber\\
    &\qquad \qquad \qquad \qquad - \hat{S}_\beta(\tau) \rI_S(t) \hat{S}_\alpha(t) \Tr_\varepsilon \Big( \hat{E}_\beta(\tau) \hat{\rho}_\varepsilon \hat{E}_\alpha(t) \Big) \nonumber\\
    &\qquad \qquad \qquad \qquad + \rI_S(t) \hat{S}_\beta(\tau) \hat{S}_\alpha(t) \Tr_\varepsilon \Big( \hat{\rho}_\varepsilon \hat{E}_\beta(\tau) \hat{E}_\alpha(t) \Big) \Big] \otimes \hat{\rho}_\varepsilon \nonumber\\
    &\eqdef - \alpha^2 \int\limits_{t_0}^t d\tau \,  \sum_{\alpha , \beta \in \mathcal{I}} \Big[ C_{\alpha \beta}(t - \tau) \Big( \hat{S}_\alpha(t) \hat{S}_\beta(\tau) \rI_S(t) - \hat{S}_\beta(\tau) \rI_S(t) \hat{S}_\alpha(t) \Big) \nonumber\\
    &\qquad \qquad \qquad \qquad + C_{\beta \alpha} (\tau - t) \Big( \rI_S(t) \hat{S}_\beta(\tau) \hat{S}_\alpha(t) - \hat{S}_\alpha(t) \rI_S(t) \hat{S}_\beta(\tau) \Big) \Big] \otimes \hat{\rho}_\varepsilon. \nonumber
\end{align}

In the last step we have defined the environmental correlation functions given by

\begin{align}
\label{eq:EnvCorr}
  C_{\alpha \beta}(t - \tau) \defeq \Tr_\varepsilon \big( \hat{E}_\alpha(t) \hat{E}_\beta(\tau) \rI_\varepsilon \big) = \big\langle \hat{E}_\alpha(t-\tau) \hat{E}_\beta(0) \big\rangle_\varepsilon.
\end{align}

The~homogeneity in the temporal argument of these functions in \eqref{eq:EnvCorr} comes from the vanishing commutator between $\hat{H}_\varepsilon$ and $\r_\varepsilon(t)$ and the cyclicity of the trace. The~latter needs to be checked explicitly depending on the type of operators in the interaction Hamiltonian. In the context of the canonical decoherence models \cite{Schlosshauer:DecoherenceBook,Breuer:OpenSystems,Schlosshauer:Boson-Spin} this assumption still holds although the environmental part of the interaction Hamiltonian is unbounded. For all of our purposes in this work we can assume that the trace is cyclic since all relevant operators will be shown to be either trace-class or bounded. Consequently, we can condense the master equation into a form involving two commutators and substitute $\xi = t-\tau$ for later~convenience:

\begin{align}
    \frac{\d}{\d t} \P \rI(t) &= - \alpha^2 \int\limits_{t_0}^t d\tau \,  \sum_{\alpha \in \mathcal{I}} \bigg( \Big[ \hat{S}_\alpha(t), \, \sum_{\beta \in \mathcal{I}} C_{\alpha \beta}(t - \tau) \hat{S}_\beta(\tau) \rI_S(t) \Big] \\
    &\qquad \qquad \qquad \quad+ \Big[ \rI_S(t) \sum_{\beta \in \mathcal{I}} C_{\beta \alpha} (\tau - t) \hat{S}_\beta(\tau), \, \hat{S}_\alpha(t) \Big] \bigg) \otimes \hat{\rho}_\varepsilon \nonumber\\
    &= - \alpha^2 \int\limits_{0}^{t-t_0} d\xi \,  \sum_{\alpha \in \mathcal{I}} \bigg( \Big[ \hat{S}_\alpha(t), \, \sum_{\beta \in \mathcal{I}} C_{\alpha \beta}(\xi) \hat{S}_\beta(t - \xi) \rI_S(t) \Big]\nonumber \\
    &\qquad \qquad \qquad \quad+ \Big[ \rI_S(t) \sum_{\beta \in \mathcal{I}} C_{\beta \alpha} (-\xi) \hat{S}_\beta(t - \xi), \, \hat{S}_\alpha(t) \Big] \bigg) \otimes \hat{\rho}_\varepsilon.\nonumber
\end{align}

At this point it is feasible to transform back into the Schrödinger picture, which essentially amounts to eliminating the time argument $t$ in the previous expressions and explicitly add the standard commutator contribution for the unitary part of the dynamics. On top of that we further abbreviate the system's operators involving the environmental correlation functions in the commutators. This yields the so-called Born--Redfield equation:

\begin{align}
    \frac{\d}{\d t} \hat{\rho}_S(t) &= -i \big[ \hat{H}_S, \hat{\rho}_S(t) \big] - \sum_{\alpha \in \mathcal{I}} \bigg( \Big[ \hat{S}_\alpha, \, \hat{B}_\alpha(t,t_0) \hat{\rho}_S(t) \Big] + \Big[ \hat{\rho}_S(t) \hat{C}_\alpha(t,t_0), \, \hat{S}_\alpha \Big] \bigg),
    \label{eq:BornRedfield}
\end{align}
where the still explicitly time-dependent operators $\hat{B}_\alpha(t,t_0)$ and $\hat{C}_\alpha(t,t_0)$ are given by:

\begin{align}
    \hat{B}_\alpha(t,t_0) &\defeq \alpha^2 \int\limits_{0}^{t-t_0} d\xi \, \sum_{\beta \in \mathcal{I}} C_{\alpha \beta}(\xi) \hat{S}_\beta(- \xi), \label{eq:B_Coeff} \\[0.3em]
    \hat{C}_\alpha(t,t_0) &\defeq \alpha^2 \int\limits_{0}^{t-t_0} d\xi \, \sum_{\beta \in \mathcal{I}} C_{\beta \alpha}(-\xi) \hat{S}_\beta(- \xi). \label{eq:C_Coeff}
\end{align}

If we take the Markov property seriously, we can further simplify Equation \eqref{eq:BornRedfield} in order to eliminate the explicit time dependence of the environmental operators $\hat{B}_\alpha(t,t_0)$ and $\hat{C}_\alpha(t,t_0)$ in \eqref{eq:B_Coeff} and \eqref{eq:C_Coeff}, respectively. Assuming that the environment has no memory effects we can conclude that the correlation functions $C_{\alpha \beta}(\xi)$ are sharply peaked (or distributional, in the exact Markovian case), which lets us extend the initial time $t_0$ to past infinity. Given that this limit exists for the model at hand, we end up with the Born-Markov master equation:

\begin{align}
    \frac{\d}{\d t} \hat{\rho}_S(t) &= -i \big[ \hat{H}_S, \hat{\rho}_S(t) \big] - \sum_{\alpha \in \mathcal{I}} \bigg( \Big[ \hat{S}_\alpha, \, \hat{B}_\alpha \hat{\rho}_S(t) \Big] + \Big[ \hat{\rho}_S(t) \hat{C}_\alpha, \, \hat{S}_\alpha \Big] \bigg),
    \label{eq:BornMarkov}
\end{align}
with the now time independent operators $\hat{B}_\alpha$ and $\hat{C}_\alpha$ given by:

\begin{align*}
    \hat{B}_\alpha \defeq \lim_{t_0 \to -\infty} \hat{B}_\alpha(t,t_0), \qquad
    \hat{C}_\alpha \defeq \lim_{t_0 \to -\infty} \hat{C}_\alpha(t,t_0).
\end{align*}

It is worth noting that, even if the limit exists, this does not guarantee that the associated master equation is of Lindblad form and hence completely positive. The~explicit outcome of these integrations crucially depends on the properties of the environmental correlation functions and the complexity of the interaction picture transformation of the system's part of the interaction Hamiltonian. Hence it is conceivable that the resulting master equation needs to be extended to achieve complete positivity, such as the Caldeira--Leggett master equation \cite{Caldeira:QBM} in the low-temperature regime \cite{Breuer:OpenSystems} in the context of quantum Brownian motion. The~minimal spin-boson model, however is naturally of Lindblad type~\cite{Schlosshauer:DecoherenceBook,Breuer:OpenSystems} and hence completely positive. We will evaluate the peakedness properties and comment on the validity of the corresponding simplifications for our proposed model in the following~sections.

\section{An Open Polymerized Scattering Model}
\label{sec:DerPolyModel}
In this section we will consider an open quantum system describing scattering in the framework of polymerized quantum mechanics. For this purpose we will in Section~\ref{sec:U1model} introduce the notation and techniques to formulate quantum mechanics on the group $U(1)$ corresponding to the situation of quantum mechanics on the unit circle. Afterwards in Section \ref{sec:U1model} we will derive the master equation of this model.

\subsection{Brief Review of Polymer Quantization and Quantum Mechanics on $U(1)$} \label{sec:U1model}
In ordinary quantum mechanics, any representation of the fundamental commutation relation of position and momentum leads to identical physical predictions by virtue of the Stone--von Neumann uniqueness result \cite{vNeumann1931eindeutigkeit}. This property relies on the (weak) continuity of the so-called Weyl elements, that is the one-parameter unitary groups associated to the position and momentum operators, which in the Schrödinger case are given by $\hat{U}(\alpha) = e^{i \alpha \hat{q}}$ and  $\hat{V}(\beta) = e^{i \beta \hat{p}}$. The~Weyl elements satisfy the following properties  

\begin{align*}
    \hat{U}(\alpha) \circ \hat{U}(\beta) = \hat{U}(\alpha + \beta), \quad
    \hat{V}(\alpha) \circ \hat{V}(\beta) = \hat{V}(\alpha + \beta),
    \quad
    \hat{U}(\alpha) \circ \hat{V}(\beta) = e^{-i\alpha \beta} \hat{V}(\beta) \circ \hat{U}(\alpha).
\end{align*}

The~task of quantization is to find a representation of the ensuing Weyl algebra on some suitable Hilbert space. The~conventional choice is simply $L^2(\mathbb{R}, \d q)$ with the standard inner product. In this representation, both Weyl elements are weakly continuous and thus the derivatives of the Weyl elements with respect to their parameters and hence the associated infinitesimal generators exist as operators on this Hilbert space.  In recent years, the so-called polymer representation has gained attention due to its inevitable presence in Loop Quantum Gravity (LQG), see for instance \cite{Thiemann:MCQGR, rovelli2004quantum,rovelli_vidotto_2014,First30YearsLQG}, and hence possible connection with Planck-scale physics. Especially in the context of Loop Quantum Cosmology (LQC), as exemplarily treated in \cite{Agullo:LQCReview,Bojowald:1999eh, Ashtekar:LQCReport,Ashtekar:MathematicalLQC,Bojowald:LQCStructure,Velhinho:2007gg,Bojowald:2005epg}, the polymer representation offers a variety of interesting aspects that are not included in the Wheeler--de Witt approach. The~crucial difference lies in the discontinuity of at least one of the  Weyl elements. The~polymer representation in one dimension can be also formulated in terms of an $L^2$ space but no longer over $\mathbb{R}$ but instead involves the so-called \textit{Bohr compactification}. In order to see this one considers the additive group $G \defeq (\mathbb{R},+)$ and uses that the corresponding characters $c \mapsto h_\mu(c) = e^{i \mu c}$ of $(\mathbb{R},+)$ labelled by $\mu \in \mathbb{R}$ form an isomorphic group that we denote by $\widetilde{G}$. Note that a locally compact Abelian group $G$ is compact if and only if $\widetilde{G}$ is discrete, so we equip $\widetilde{G}$ with the discrete topology, where this brief introduction closely follows the notation in \cite{Fewster:2008sr}. Consequently, the dual of $\widetilde{G}$, more precisely its set of characters, is the afore-mentioned Bohr compactification $G_b$ of the original group $G$, which carries a natural probability measure in terms of the Haar measure $\d \mu_H$ \cite{Ashtekar:Polymer,corichi2007polymer,Fewster:2008sr}. The~polymer Hilbert space $\mathcal{H}_{\textrm{poly}}$ and its most convenient orthonormal basis are then given by

\begin{align*}
    \mathcal{H}_{\textrm{poly}} \defeq L^2(\mathbb{R}_b, \d \mu_H), \quad e_\mu(q) \defeq e^{i \mu q}\quad{\rm  with}\quad  \mu \in \mathbb{R}, 
\end{align*}
where the basis is given in \textit{q-polarization}.
In this representation, that we call the A-type representation following the notation from  \cite{corichi2007polymer}, the fundamental operators act upon the basis elements according to

\begin{align}
    \hat{U}(\alpha) \ket{\mu} = \ket{\mu + \alpha}\quad{\rm  and}\quad \hat{p}\ket{\mu} = \mu \ket{\mu}.
\end{align} 

The~inner product is defined via the Haar measure on $\mathbb{R}_b$:

\begin{align*}
   \big\langle e_\mu, e_\nu \big\rangle_{\textrm{poly}} = \int\limits_{\mathbb{R}_b} \d \mu_H \; \bar{e}_\mu(q) e_\nu(q) \defeq \lim_{L \to \infty} \frac{1}{2L} \int\limits_{-L}^L \d q \; e^{i (\nu - \mu) q} = \delta_{\mu, \nu}.
\end{align*}

Due to the uncountably infinite number of basis elements, $\mathcal{H}_{\textrm{poly}}$ is non-separable. Furthermore, in constrast to the Schrödinger representation the Weyl element $\hat{U}(\alpha)$ is not weakly continuous at $\alpha = 0$ since any two states $\ket{\mu}, \ket{\mu + \varepsilon}$ are orthogonal for all $\varepsilon \in \mathbb{R}$ with $|\varepsilon| > 0$. Hence, the position operator as the generator of $\hat{U}(\alpha)$ does not exist on $\mathcal{H}_{\textrm{poly}}$. The~momentum operator $\hat{p}$, however, exists based on the weak continuity of $\hat{V}(\beta)$. By means of an Application of the Fourier--Bohr transform of the basis elements denoted by $ \mathcal{F}[e_\mu](p)$ we obtain the dual basis of the Hilbert space $L^2(\mathbb{R}_d, \d \mu_c)$ with $\mathbb{R}_d$ denoting the real numbers with discrete topology and $\d \mu_c$ the associated counting measure

\begin{align*}
    \mathcal{F}[e_\mu](p) = \int\limits_{\mathbb{R}_b} \d \mu_H \; \bar{e}_\mu(q) e^{-ipq} \defeq \lim_{L \to \infty} \frac{1}{2L} \int\limits_{-L}^L \d q \; e^{i (p - \mu) q} = \delta_{\mu, p} \eqdef \tilde{e}_\mu(p).
\end{align*}

The~dual basis $\tilde{e}_\mu(p) = \delta_{\mu, p}$ is orthonormal with respect to the counting measure, that~is

\begin{align*}
    \big\langle \tilde{e}_\mu, \tilde{e}_\nu \big\rangle_{\textrm{poly}} = \int\limits_{\mathbb{R}_d} \d \mu_c \; \bar{\tilde{e}}_\mu(p) \tilde{e}_\nu(p) \defeq \sum_{p \in \mathbb{R}} \delta_{\mu, p} \delta_{\nu, p} = \delta_{\mu, \nu}.
\end{align*}

Likewise, the basis $\delta_{\mu, p}$ is uncountable and $L^2(\mathbb{R}_d, \d \mu_c)$ non-separable as a consequence. It can be shown that he polymer Hilbert space $\mathcal{H}_{\textrm{poly}}$ decomposes into a direct sum of separable Hilbert spaces \cite{Ashtekar:Polymer}. We have

\begin{equation}
    \mathcal{H}_{\textrm{poly}} = \bigoplus_{\delta \in [0,1)} \mathcal{H}_{\textrm{poly}}^{(\delta)}, \quad \big\langle e_n^{(\delta)}, e_m^{(\delta)} \big\rangle_{\textrm{poly}}^{(\delta)} \defeq \int\limits_{-\pi}^\pi \frac{\d \varphi}{2 \pi} \; \bar{e}_n^{(\delta)} e_m^{(\delta)} = \delta_{n,m}, \label{eq:supersec_product}
\end{equation}
where $\mathcal{H}_{\textrm{poly}}^{(\delta)}$ are so called superselection Hilbert spaces and their basis elements $e_n^{(\delta)}$ are obtained by realizing that the basis label $\mu \in \mathbb{R}$ can be decomposed into an integer $n$ and a superselection parameter $\delta$ in the following~manner. Different aspects on how this parameter arises from representation theory on $L^2(S^1, \d \varphi / 2\pi)$ and the physical implications of different values of $\delta$ in the context of various physical phenomena are briefly reviewed in Appendix \ref{sec:bead_circle}.

\begin{align}
    e_\mu(q) = e^{i\mu q} = e^{i(n+\delta)q} \eqdef e_n^{(\delta)}(q), \quad n \in \mathbb{Z}, \quad \delta \in [0,1) \subset \mathbb{R}, \label{eq:supersec_basis}
\end{align}

The~term \textit{superselection} refers to the fact that the fundamental operators $\hat{U}(\alpha)$ and $\hat{p}$ that generate the algebra of observables do not map between individual superselection sectors if $\alpha$ is constrained to integer values. These superselection Hilbert spaces carry an inner product akin to the ordinary choice for the square-integrable functions on a unit circle, $L^2(S^1, \d \varphi / 2\pi, \delta)$, where $\delta$ is associated to the quasi-periodicity of basis functions. This~freedom in the choice of representation is also briefly discussed in Appendix \ref{sec:bead_circle}. 
At~this point we adopt the conventional notation also used in \cite{Kastrup:AnglePair,Bahr:Coherent} in terms of $\varphi$ as the angle variable and $p_\varphi$ as its associated (angular) momentum. A connection to the notation often used in the literature for polymerized quantum systems can be established by rescaling

\begin{align*}
    n \mapsto n\mu_0,\quad \delta \mapsto \mu_0 \delta=:p_0 \quad {\rm and}\quad q \mapsto \mu_0^{-1} q,
\end{align*}
and a corresponding rescaling of the limits in the inner product. In this case $\mu_0$ can be associated to some fundamental length scale $\mu_0$ originating from a more profound theory, providing the lattice spacing of the discrete momentum variable. If we were to polymerize the translation generator $\hat{V}$, this fundamental discreteness would be found in the eigenvalues of the position operator, analogous to, but much simpler than, the occurrence of an area gap in LQG \cite{Thiemann:MCQGR,rovelli2004quantum} and the upper bound on matter density in LQC \cite{Ashtekar:LQCReport,Ashtekar:2007robustness} respectively. However, the rescaling in terms of integer multiples of $\mu_0$ does not add any additional insights and can be shown to lead to the same \textit{qualitative} results in terms of the structural properties of the master equation. Note however, that the polymer scale explicitly enters the master equation in terms of the eigenvalues of the momentum operator, hence we do \textit{not} observe scale invariance in the model. The~elementary holonomy and momentum operators in the model considered here act on the basis elements given in (\ref{eq:supersec_basis}):

\begin{align*}
 \hat{U}(\lambda) \ket{n}_\delta = \ket{n+\lambda}_\delta,\quad \hat{U}(\lambda)^\dagger \ket{n}_\delta = \ket{n-\lambda}_\delta   \quad {\rm and}\quad
 \hat{p}_\varphi \ket{n}_\delta = \big( n+\delta \big) \ket{n}_\delta.
\end{align*}

For the holonomy operators we directly see that $\lambda \notin \mathbb{Z}$ would map out of  a given superselection sector.  All further calculations will be performed in the $e_n^{(\delta)}$ basis states shown in (\ref{eq:supersec_basis}), which are the momentum eigenstates. In the next subsection we will use the individual superselection sectors in order to derive a quantum master equation for an open $U(1)$ scattering model for which we will specify the corresponding interaction~Hamiltonian.

\subsection{Derivation of the Master Equation of an Open $U(1)$ Scattering Model} \label{sec:model_application}
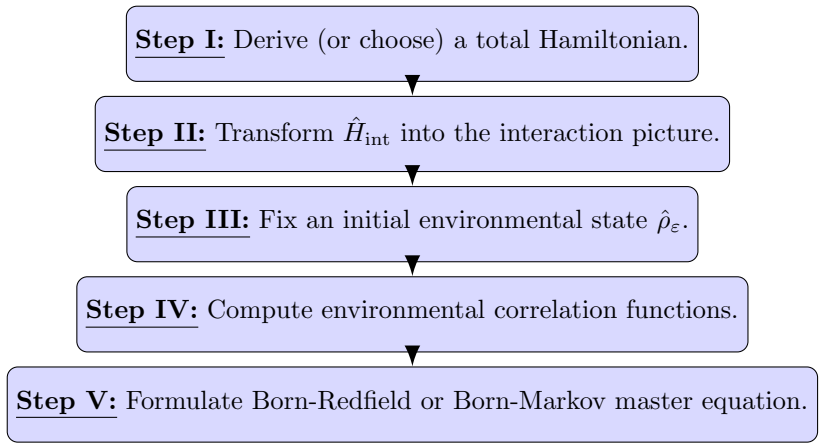
\begin{figure}[H] \centering
		\begin{tikzpicture}[node distance=6cm]
		\node 	(pro2) [process]  {\begin{varwidth}{\linewidth} 
			\centering \underline{\textbf{Step I:}}
			Derive (or choose) a total Hamiltonian. \end{varwidth}};
    	\node(pro3) [process, below of=pro2, node distance = 1.2cm] 
			{\begin{varwidth}{\linewidth}
			\centering \underline{\textbf{Step II:}}
			Transform $\hat{H}_{\textrm{int}}$ into the interaction picture. \end{varwidth}};
		\node 	(pro4) [process, below of=pro3, node distance = 1.2cm]                {\begin{varwidth}{\linewidth} 
			\centering \underline{\textbf{Step III:}}
			Fix an initial environmental state $\hat{\rho}_\varepsilon$. \end{varwidth}};
		\node 	(pro5) [process, below of=pro4, node distance = 1.2cm]                {\begin{varwidth}{\linewidth} 
			\centering \underline{\textbf{Step IV:}}
			Compute environmental correlation functions. \end{varwidth}};
		\node 	(pro6) [process, below of=pro5, node distance = 1.2cm]                {\begin{varwidth}{\linewidth} 
			\centering \underline{\textbf{Step V:}}
			Formulate Born-Redfield or Born-Markov master equation. \end{varwidth}};
	\draw[arrow](pro2.south) to [out=90,in=90] node[below]{} (pro3.north);
	\draw[arrow](pro3.south) to [out=90,in=90] node[below]{} (pro4.north);
	\draw[arrow](pro4.south) to [out=90,in=90] node[below]{} (pro5.north);
	\draw[arrow](pro5.south) to [out=90,in=90] node[below]{} (pro6.north);
    \end{tikzpicture} \vspace*{0.5cm}
    \caption{Illustration of the major steps in the derivation of a master equation.} \label{fig1}
\end{figure}

A minimal decoherence model unarguably consists of free particles in system and environment with a given interaction Hamiltonian \textit{bilinear} in the combined variables. In our case, the analogue of the position operator is unitary and hence bounded, contrary to standard quantization procedures, a fact which at some points during the derivation of the master equation will turn out to be beneficial regarding the appearance of singularities in a straightforward approach, as was already briefly discussed in \cite{hornberger2003collisional}. In order to feature a common thread in the several steps of the derivation of a master equation, we would like to give the reader a concise overview in the form of the following flowchart in Figure \ref{fig1}: \\ 

In terms of the A-representation in which $U(\lambda)$ is discontinuous and the position (or angle) operator does hence not exist, we associate a coupling constant $g_\alpha$ to each individual particle in the environment. This is a technicality related to the physical dimension of the coupling constant. In this model, the coupling constant is not dimensionless, since the interaction Hamiltonian has a fixed dimension by definition. In the course of the perturbative expansion, the strategy is completely analogous to the dimensionless case. At the level of the master equation, this choice of interaction enables us in principle to introduce a spectral density in order to emulate a continuum of (effective) particle masses in the environment, which in some cases is a crucial ingredient in the derivation of a Lindblad-type master equation. An interaction Hamiltonian associated to a dynamic position monitoring \cite{Breuer:OpenSystems,Schlosshauer:DecoherenceBook} or \textit{localization} includes a coupling of the position variables of system and environment, respectively, hence our \textit{choice} of total Hamiltonian is given by:

\begin{align}
    \hat{H}_{\textrm{tot}} &= \hat{H}_S \otimes \mathds{1}_\varepsilon + \mathds{1}_S \otimes \hat{H}_\varepsilon + \hat{H}_{\textrm{int}}, \qquad \textrm{with} \qquad \hat{H}_{\textrm{int}} \defeq \sum_{\alpha = 1}^N \hat{S} \otimes \hat{E}_\alpha, \\[0.4em]
    &= \frac{\hat{p}_\varphi^2}{2 m_0} \otimes \mathds{1}_\varepsilon + \mathds{1}_S \otimes \sum_{\alpha = 1}^N \frac{\big( \hat{p}_\varphi^{(\alpha)} \big)^2}{2 m_\alpha} + \frac{1}{2i\lambda} \Big( \hat{U}_\lambda - \hat{U}_\lambda^\dagger \Big) \otimes \sum_{\alpha = 1}^N  \frac{g_\alpha}{2 i \lambda} \Big( \hat{U}^{(\alpha)}_\lambda - \big( \hat{U}^{(\alpha)}_\lambda \big)^\dagger \Big). \label{eq:tot_hamiltonian}
\end{align}
where the index $\alpha$ corresponds to the individual particles in the (thermal) environment, $g_\alpha$ is the respective coupling and $N \gg 1$ is the particle number. The~assumption that $N$ is large is based on a technicality that relates to the thermodynamic limit, more precisely to the replacement of the sum over couplings by an integral over a spectral density. This will become clear in the process of deriving and discussing the final master equation. Furthermore, the notation $\hat{U}^{(\alpha)}_\lambda \defeq \hat{U}_\alpha(\lambda)$ was adopted to keep distinguishability between the Schrödinger- and the interaction picture operators. The~prefactors of $(2i\lambda)^{-1}$ for every such combination of holonomies ensure that the Hamiltonian has an appropriate limit for vanishing $\lambda$. The~connection to similar models in a Schrödinger-like representation \cite{Breuer:OpenSystems,Schlosshauer:DecoherenceBook,Caldeira:QBM} is given by the replacement:

\begin{align}
    q \otimes \sum_{\alpha = 1}^N g_\alpha q_\alpha \to \frac{1}{2i\lambda} \Big( \hat{U}_\lambda - \hat{U}_\lambda^\dagger \Big) \otimes \sum_{\alpha = 1}^N  \frac{g_\alpha}{2 i \lambda} \Big( \hat{U}^{(\alpha)}_\lambda - \big( \hat{U}^{(\alpha)}_\lambda \big)^\dagger \Big).
\end{align}

In this sense, $\lambda$ describes a lattice spacing where $\lambda = 1$ is the non-rescaled case described in Appendix \ref{sec:bead_circle}. However intermediate values are not meaningful in the sense that they would map out of the chosen superselection sector. This procedure of emulating operators via differences of holonomies is well-known in the literature, see for example~\cite{Ashtekar:Polymer}, but has never been applied to the canonical open quantum mechanical decoherence models~\cite{Caldeira:QBM,Schlosshauer:Boson-Spin,Breuer:OpenSystems} to the knowledge of the authors. On top of this, for the model considered here, the continuum limit might not be well-defined in the midst of our calculations. This is not surprising since there are divergences associated to the fact that momentum eigenstates are non-normalizable in Schrödinger quantum mechanics, see for instance the discussion in \cite{hornberger2003collisional}. This can however be in principle dealt with in the framework of rigged Hilbert spaces \cite{de2005role}. In the model considered here these kind of divergencies are absent because we work in the polymer but not in the Schrödinger representation and as a consequence the position variable $q$  needs to be expressed in terms of holonomies, often referred to as the polymerization of the position variable. The~total Hilbert space has the following structure:

\begin{align}
    \mathcal{H}_{\textrm{tot}} &= \mathcal{H}_S \otimes \mathcal{H}_\varepsilon = L^2\Big(S^1, \frac{\d \varphi}{2\pi}, \delta\Big) \otimes \Bigg[ \bigotimes_{\alpha = 1}^N L^2\Big( S^1_{(\alpha)}, \frac{\d \varphi^{(\alpha)}}{2\pi}, \delta \Big) \Bigg].
\end{align}

Note that the superselection parameter $\delta$ agrees for all environmental degrees of freedom, since adding a particle label to the parameter does add notational complexity while leaving the results largely unchanged. Considering the continuum limit in terms of a spectral density would be a more complicated affair, however. Given a total Hamiltonian, the next step towards a master equation of the form (\ref{eq:BornRedfield}) is to transform the interaction Hamiltonian into the interaction picture. The~most straightforward route is to solve the Heisenberg equations of motion for the operators involved in the interaction:

\begin{equation}
    \frac{\d}{\d \xi} \hat{U}_\lambda(\xi) = i \big[\hat{H}_{\textrm{tot}}, \hat{U}_\lambda(\xi) \big] + \bigg( \frac{\partial}{\partial \xi} \hat{U}_\lambda \bigg)(\xi),
\end{equation}
where an explicit time argument denotes the interaction picture quantity, with $\hat{U}_\lambda \eqdef \hat{U}_\lambda(0)$ being the Schrödinger picture operator. Due to the same structure of $\hat{H}_S$ and $\hat{H}_\varepsilon$ respectively, it is sufficient to investigate the interaction picture of either part of the interaction Hamiltonian, the other one follows immediately. The~fact that $[\hat{U}_\lambda, \hat{p}_\varphi] = -\lambda \hat{U}_\lambda$ and $[\hat{U}^\dagger_\lambda, \hat{p}_\varphi] = \lambda \hat{U}^\dagger_\lambda$ together with the time-independence of $\hat{U}_\lambda$ and $\hat{U}^\dagger_\lambda$ lets us conclude:

\begin{align*}
    \frac{\d}{\d \xi} \hat{U}_\lambda(\xi) &= \frac{i}{2m_0} e^{ i \xi \hat{H}_{S}} \big[\hat{p}_\varphi^2, \hat{U}_\lambda \big] e^{-i \xi \hat{H}_{S}} = \frac{i\lambda}{m_0} \hat{U}_\lambda(\xi) \bigg( \hat{p}_\varphi + \frac{\lambda \mathds{1}}{2} \bigg) = \frac{i\lambda}{m_0} \bigg(\hat{p}_\varphi - \frac{\lambda \mathds{1}}{2} \bigg) \hat{U}_\lambda(\xi), \\[0.4em]
    \frac{\d}{\d \xi} \hat{U}^\dagger_\lambda(\xi) &= \frac{i}{2m_0} e^{ i \xi \hat{H}_{S}} \big[\hat{p}_\varphi^2, \hat{U}^\dagger_\lambda \big] e^{-i \xi \hat{H}_{S}} = \frac{i\lambda}{m_0} \hat{U}^\dagger_\lambda(\xi) \bigg( - \hat{p}_\varphi + \frac{\lambda \mathds{1}}{2} \bigg) = - \frac{i\lambda}{m_0} \bigg(\hat{p}_\varphi + \frac{\lambda \mathds{1}}{2} \bigg) \hat{U}^\dagger_\lambda(\xi).
\end{align*}

Hence we can directly give a closed-form expression in terms of an exponential: 

\begin{align}
   \hat{U}_\lambda(\xi) &= \hat{U}_\lambda(0) \exp{\frac{i \lambda \xi}{m_0} \bigg( \hat{p}_\varphi + \frac{\lambda \mathds{1}}{2} \bigg)} = \exp{\frac{i \lambda \xi}{m_0} \bigg( \hat{p}_\varphi - \frac{\lambda \mathds{1}}{2} \bigg)} \hat{U}_\lambda(0), \label{eq:int_picture_U} \\[0.4em]
   \hat{U}^\dagger_\lambda(\xi) &= \hat{U}^\dagger_\lambda(0) \exp{\frac{i \lambda \xi}{m_0} \bigg( - \hat{p}_\varphi + \frac{\lambda \mathds{1}}{2} \bigg)} = \exp{- \frac{i \lambda \xi}{m_0} \bigg( \hat{p}_\varphi + \frac{\lambda \mathds{1}}{2} \bigg)} \hat{U}^\dagger_\lambda(0). \label{eq:int_picture_Udagger}
\end{align}

The~fact that the above forms of (\ref{eq:int_picture_U}) and (\ref{eq:int_picture_Udagger}) are the correct solutions can be explicitly checked by direct computation of the derivative or by virtue of 'traditionally' evaluating the iterated commutator bracket, see Appendix \ref{sec:appendixB}. The~next step consists of the computation of the environmental two-point correlation functions, where we have chosen a thermal state for the environment, which is the most convenient choice and has the benefit of time-homogeneous correlation functions, that is $C_{\alpha \beta}(t,\tau) = C_{\alpha \beta}(t-\tau)$. Thus we need to evaluate $C_{\alpha \beta}(\xi) = \big\langle\hat{E}_\alpha(\xi) \hat{E}_\beta(0) \big\rangle_\varepsilon$, where $\big\langle \cdot \big\rangle_\varepsilon$ denotes the environmental expectation value with an $N$-particle thermal environment $\r_\varepsilon = \bigotimes_i \rS$. For this purpose we use the framework introduced in Section \ref{sec:U1model} where we can construct a thermal state that can act as an $N$-particle environment for our single system degree of freedom. As usual, we start out with a thermally weighted exponential of the respective free particle Hamiltonian $e^{-\beta \hat{H}}$ with $\beta = (k_B T)^{-1}$ where $k_B$ is the Boltzmann constant and $T$ is the temperature parameter. Consider a single particle in a thermal state, which we can then easily extend to $N$ identical particles for the environment and hence skip the particle label $\alpha$ for now. First expand the Hamiltonian in the basis of (angular) momentum eigenstates:

\begin{align*}
    \rS &\defeq \frac{e^{-\beta \hat{H}}}{\Tr{e^{-\beta \hat{H}}}}
    = \frac{1}{Z_{\rm{th}}} \sum_{n \in \mathbb{Z}} \exp{-\frac{\beta}{2m} \big( n+\delta \big)^2} \ket{n}_\delta \otimes \bra{n}_\delta \\[0.4em]
    &= \frac{e^{-\sigma \delta^2}}{Z_{\rm{th}}} \sum_{n \in \mathbb{Z}} \exp{-\sigma n^2 - 2 \sigma n \delta} \ket{n}_\delta \otimes \bra{n}_\delta, \quad \sigma \defeq \frac{\beta}{2m}.
\end{align*}

\begin{align*}
    Z_{\rm{th}} \defeq \Tr{e^{-\beta \hat{H}}} &= \sum_{m,n \in \mathbb{Z}} \exp{-\sigma n^2 - 2 \sigma n \delta - \sigma \delta^2} \bra{m} \ket{n}_\delta \bra{n} \ket{m}_\delta \\[0.4em]
    &= \sum_{n \in \mathbb{Z}} \exp{- \sigma \delta^2} \exp{-\sigma n^2 - 2 \sigma n \delta} \\[0.4em]
    &= \exp{- \sigma \delta^2} \sum_{n \in \mathbb{Z}} \exp{i \pi n^2 \tau + 2 \pi i n z} \\[0.3em]
    &\eqdef \exp{- \sigma \delta^2} \vartheta(z;\tau),
\end{align*}
where $\vartheta(z;\tau)$ (or often $\vartheta_{00}(z;\tau)$ if considered in the context of auxiliary theta functions) is the Jacobi theta function, $z=i \sigma \delta / \pi$ and $\tau = i \sigma / \pi$ for brevity. Note that $\vartheta(z;\tau)$ is non-vanishing and real for all $\sigma$ regardless of the temperature parameter, hence the density operator for a single free particle on a circle in a thermal state with temperature $T$ is given by the expression:

\begin{equation}
    \rS = \frac{1}{\vartheta(z;\tau)}  \sum_{n \in \mathbb{Z}} \exp{-\sigma n^2 - 2 \sigma n \delta} \ket{n}_\delta \otimes \bra{n}_\delta.
    \label{eq:single_thermal}
\end{equation}

In the limit of $T \to 0$ or $\mathfrak{Im}(\tau) \to \infty$, respectively, we can easily see that $\vartheta(z;\tau) \to 1$ and the density operator reduces to the vacuum state, that is $\rS \to \ket{0}_\delta \otimes \bra{0}_\delta$ as in standard quantum mechanics.  Since any given $\hat{U}^{(\alpha)}_\lambda$ (note that we have omitted the subspace index in the previous calculations for a better overview) only acts on the $\alpha$-subspace of the environmental Hilbert space and maps to a state orthogonal to the previous one, we can directly conclude that $C_{\alpha \beta}(\xi) \sim \delta_{\alpha \beta}$. This of course is nothing but another justification of one of the steps we took in the course of the derivation of the master equation. It amounts to the statement that $\P \mathcal{L}(t_1) \ldots \mathcal{L}(t_{2n+1}) \P \rI(t) = 0$, in particular for the case $n=0$, since operators with different indices act on different environmental subspaces. Thus it is sufficient to find $C_{\alpha \alpha}(\xi)$ according to the following, where contributions including two instances of $\hat{U}^{(\alpha)}_\lambda$ or $\big( \hat{U}^{(\alpha)}_\lambda \big)^\dagger$ can be dropped due to orthogonality of the $\ket{m} = \ket{m_1} \otimes \ldots \otimes \ket{m_N}$ basis states:

\begin{align*}
    C_{\alpha \alpha}(\xi) &= \sum_{\{m_\iota\}} \bra{m_1, \ldots m_N} \hat{E}_\alpha(\xi) \hat{E}_\alpha(0) \Big[ \sum_{\{n_\iota\}} \bigotimes_{\iota=1}^N \frac{\exp\big\{ -\sigma_\iota n_\iota^2 - 2 \sigma_\iota n_\iota \delta \big\}}{\vartheta(z_\iota,\tau_\iota)} \ket{n_\iota}_\delta \otimes \bra{n_\iota}_\delta \Big] \ket{m_1, \ldots m_N} \\[0.2em]
    &= \sum_{\{n_\iota\}}^{\iota \neq \alpha} \prod_{\stackrel{\iota=1}{\iota \neq \alpha}}^{N} \frac{\exp\big\{ -\sigma_\iota n_\iota^2 - 2 \sigma_\iota n_\iota \delta \big\}}{\vartheta(z_\iota,\tau_\iota)} \sum_{n_\alpha \in \mathbb{Z}} \frac{\exp\big\{ -\sigma_\alpha n_\alpha^2 - 2 \sigma_\alpha n_\alpha \delta \big\}}{\vartheta(z_\alpha,\tau_\alpha)} \bra{n_\alpha} \hat{E}_\alpha(\xi) \hat{E}_\alpha(0) \ket{n_\alpha}_\delta \\
    &=  \sum_{n_1=1}^{N} \frac{\exp\big\{ -\sigma_1 n_1^2 - 2 \sigma_1 n_1 \delta \big\}}{\vartheta(z_1,\tau_1)} \times \ldots \times \sum_{n_\alpha \in \mathbb{Z}} \frac{\exp\big\{ -\sigma_\alpha n_\alpha^2 - 2 \sigma_\alpha n_\alpha \delta \big\}}{\vartheta(z_\alpha,\tau_\alpha)} \bra{n_\alpha} \hat{E}_\alpha(\xi) \hat{E}_\alpha(0) \ket{n_\alpha}_\delta \\[0.2em]
    &= \frac{g^2_\alpha}{4 \lambda^2} \sum_{n \in \mathbb{Z}} \frac{\exp\big\{ -\sigma_\alpha n^2 - 2 \sigma_\alpha n \delta \big\}}{\vartheta(z_\alpha,\tau_\alpha)} \bra{n} \hat{U}^{(\alpha)}_\lambda(\xi) \big( \hat{U}^{(\alpha)}_\lambda \big)^\dagger + \big( \hat{U}^{(\alpha)}_\lambda \big)^\dagger(\xi) \hat{U}^{(\alpha)}_\lambda \ket{n}_\delta \\[0.2em]
    &= \frac{g^2_\alpha}{4 \lambda^2} \sum_{n \in \mathbb{Z}} c_n^{(\alpha)} \bigg[ \exp\Big\{ - \frac{i \xi}{2 m_\alpha} \Big( \lambda^2 - 2 \lambda \big( n + \delta \big) \Big) \Big\} + \exp\Big\{ - \frac{i \xi}{2 m_\alpha} \Big( \lambda^2 + 2 \lambda \big( n + \delta \big) \Big) \Big\} \bigg] \\
    &= \frac{g^2_\alpha}{2 \lambda^2} \sum_{n \in \mathbb{Z}} c_n^{(\alpha)} \bigg[ \cos\bigg(\frac{\xi \lambda^2}{2 m_\alpha}\bigg) - i \sin\bigg(\frac{\xi \lambda^2}{2 m_\alpha} \bigg) \bigg] \cos\bigg(\frac{\xi \lambda}{m_\alpha}\big( n + \delta \big)\bigg) \\[0.2em]
    &\eqdef \nu_\alpha(\xi) - i \eta_\alpha(\xi),
\end{align*}

where in the first step, we decomposed the $N$-particle Hilbert space scalar product into its individual contributions. In the second step we isolated the only non-trivial contribution to this product, note that $\hat{E}_\alpha(\xi) \hat{E}_\alpha(0)$ only acts on the respective subspace, all other elements can be pulled through, where we then use orthonormality of the single-particle basis states. This gives the definitive proof that only correlation functions with $\alpha = \beta$ give a non-vanishing contribution, $\bra{n_\alpha} \hat{E}_\alpha(\xi) \ket{n_\alpha}_\delta \sim \bra{n_\alpha} \ket{n_\alpha+\lambda}_\delta + \bra{n_\alpha} \ket{n_\alpha-\lambda}_\delta = 0$. Each individual non-$\alpha$ instance of the contributions including the Jacobi theta function sum up to unity, simply due to the form of the partition function, see Equation (\ref{eq:single_thermal}). Consequently, we abbreviated the normalization factor as $c_n^{(\alpha)}$ and renamed the summation index from $n_\alpha$ to $n$ but  kept this label on the coefficients, as for example $\tilde{\beta}_\alpha$ explicitly contains the mass $m_\alpha$. This parameter, which we did not yet a priori fix among the environmental degrees of freedom, exhibits an important aspect and will be brought up again later in this work. In the last step we just applied the very definitions of $\hat{U}^{(\alpha)}_\lambda(\xi)$ and $\big( \hat{U}^{(\alpha)}_\lambda \big)^\dagger$, respectively and abbreviated the result in terms of the so-called \textit{noise-} and \textit{dissipation} kernel functions $\nu_\alpha(\xi)$ and $\eta_\alpha(\xi)$ respectively. The~origin of this nomenclature will be discussed in the next section, where we will shed light onto the fact that this straightforward interpretation breaks down amidst the polymerization of the configuration variable. Note that $C_{\alpha \alpha}(-\xi) = \big\langle\hat{E}_\alpha(0) \hat{E}_\alpha(\xi) \big\rangle_\varepsilon = \big\langle\hat{E}_\alpha(-\xi) \hat{E}_\alpha(0) \big\rangle_\varepsilon$ can be checked to hold explicitly for our particular correlation functions. The~fact that there is only a single system's degree of freedom and $C_{\alpha \beta}(-\xi) \sim \delta_{\alpha \beta}$ eases our way towards the final result.

At this stage we can directly insert the correlation functions $C_{\alpha \alpha}(-\xi)$ into the Born--Redfield master Equation (\ref{eq:BornRedfield}). We intend to keep the explicit time-dependence from the upper limit of integration as long as possible, taking the limit $t_0 \to \infty$ is generally non-trivial and only possible for certain models and spectral densities, the latter will be discussed in more detail in Section \ref{sec:model_interpretation}. Given the concrete form of the environmental correlation functions and depending on the temperature of the environment, it might be reasonable to introduce another aspect of the Markov approximation and extend the integral limits accordingly, but for now we shall remain in Redfield form. For the benefit of the reader we recall the individual contributions in the Redfield master equation:

\begin{align*}
    \frac{\d}{\d t} \hat{\rho}_S(t) &= -i \big[ \hat{H}_S, \hat{\rho}_S(t) \big] - \sum_{\alpha \in \mathcal{I}} \bigg( \Big[ \hat{S}_\alpha, \, \hat{B}_\alpha(t,t_0) \hat{\rho}_S(t) \Big] + \Big[ \hat{\rho}_S(t) \hat{C}_\alpha(t,t_0), \, \hat{S}_\alpha \Big] \bigg).
\end{align*}

In the light of the definitions in (\ref{eq:B_Coeff}) and (\ref{eq:C_Coeff}) together with the realization that the sum over $\alpha$ only affects the environmental contributions, for the system consists only of a single degree of freedom, we can further simplify.  Utilizing that the interaction picture operators of the system's part of the interaction Hamiltonian are given in full analogy to the environment, we get:

\begin{align*}
    \hat{U}_\lambda(-\xi) = \hat{U}_\lambda \exp\Big\{ -\frac{i \xi \lambda}{m_0} \Big( \hat{p}_\varphi + \frac{\lambda \mathds{1}}{2} \Big) \Big\}, \qquad
    \hat{U}_\lambda^\dagger (-\xi) = \hat{U}_\lambda^\dagger \exp\Big\{ \frac{i \xi \lambda}{m_0} \Big( \hat{p}_\varphi - \frac{\lambda \mathds{1}}{2} \Big) \Big\}.
\end{align*}

As can be checked explicitly, it holds that $\big( \hat{U}_\lambda(\xi) \big)^\dagger = \big( \hat{U}_\lambda\big)^\dagger(\xi)$. This can be easily proven by adjoining the previous result and with the help of the Weyl algebra relations. Straightforward insertion into the definitions of (\ref{eq:B_Coeff}) and (\ref{eq:C_Coeff}) then yields the environmental monitoring operators for the open $U(1)$ scattering model:

\begin{align*}
    \hat{B}_\alpha(t,t_0) &= \frac{1}{2 i \lambda} \int\limits_{0}^{t-t_0} d\xi \,  C_{\alpha \alpha}(\xi) \Big( \hat{U}_\lambda(-\xi) - \hat{U}_\lambda^\dagger (-\xi) \Big) \\
    &= \frac{g^2_\alpha}{4 i \lambda^3} \int\limits_{0}^{t-t_0} d\xi \, \sum_{n \in \mathbb{Z}} c_n^{(\alpha)} \bigg[ \cos\bigg(\frac{\xi \lambda^2}{2 m_\alpha}\bigg) - i \sin\bigg(\frac{\xi \lambda^2}{2 m_\alpha} \bigg) \bigg] \cos\bigg(\frac{\xi \lambda}{m_\alpha}\big( n + \delta \big)\bigg) \\
    &\times \bigg( \hat{U}_\lambda \exp\Big\{ -\frac{i \xi \lambda}{m_0} \Big( \hat{p}_\varphi + \frac{\lambda \mathds{1}}{2} \Big) \Big\} - \hat{U}_\lambda^\dagger \exp\Big\{ \frac{i \xi \lambda}{m_0} \Big( \hat{p}_\varphi - \frac{\lambda \mathds{1}}{2} \Big) \Big\} \bigg),
\end{align*}
and similarly for the second set of operators:

\begin{align*}
    \hat{C}_\alpha(t,t_0) &= \frac{1}{2 i \lambda} \int\limits_{0}^{t-t_0} d\xi \,  C_{\alpha \alpha}(-\xi) \Big( \hat{U}_\lambda(-\xi) - \hat{U}_\lambda^\dagger (-\xi) \Big) \\
    &= \frac{g^2_\alpha}{4 i \lambda^3} \int\limits_{0}^{t-t_0} d\xi \, \sum_{n \in \mathbb{Z}} c_n^{(\alpha)} \bigg[ \cos\bigg(\frac{\xi \lambda^2}{2 m_\alpha}\bigg) + i \sin\bigg(\frac{\xi \lambda^2}{2 m_\alpha} \bigg) \bigg] \cos\bigg(\frac{\xi \lambda}{m_\alpha}\big( n + \delta \big)\bigg) \\
    &\times \bigg( \hat{U}_\lambda \exp\Big\{ -\frac{i \xi \lambda}{m_0} \Big( \hat{p}_\varphi + \frac{\lambda \mathds{1}}{2} \Big) \Big\} - \hat{U}_\lambda^\dagger \exp\Big\{ \frac{i \xi \lambda}{m_0} \Big( \hat{p}_\varphi - \frac{\lambda \mathds{1}}{2} \Big) \Big\} \bigg).
\end{align*}

Furthermore we can set $\hat{B}(t,t_0) \defeq \sum_\alpha \hat{B}_\alpha(t,t_0)$ and $\hat{C}(t,t_0) \defeq \sum_\alpha \hat{C}_\alpha(t,t_0)$, respectively. As mentioned, this is specifically possible in the model here since the summation indices between the system's interaction picture operators $\hat{S}_\beta(-\xi)$ and the environmental correlation functions $C_{\alpha \beta}(-\xi)$ are not intertwined due to the fact that the $C_{\alpha \beta}(-\xi)$ are diagonal and the system only has a single degree of freedom. For a brief moment we will resert to the more compact notation $\hat{U}_\lambda (-\xi)$ and $\hat{U}_\lambda^\dagger (-\xi)$ instead of the above form. This lets us see the structure of the resulting master equation more clearly, we will reconsider the explicit, written-out form when we examine the physical properties of the individual contributions in Section \ref{sec:model_interpretation}. Let us briefly recall the abbreviations we made along the way:

\begin{align*}
    \vartheta(z_\alpha, \tau_\alpha) \defeq \sum_{n \in \mathbb{Z}} e^{i \pi n^2 \tau_\alpha} e^{2i \pi n z_\alpha}, \quad z_\alpha = \frac{i \sigma_\alpha \delta}{\pi}, \quad \tau_\alpha = \frac{i \sigma_\alpha}{\pi}, \quad \sigma_\alpha \defeq \frac{\beta}{2m_\alpha} = \frac{1}{2 m_\alpha k_B T}.
\end{align*}

Gathering all individual terms we can finally insert these into the Born-Redfield Equation (\ref{eq:BornRedfield}) and obtain:

\begin{align*}
    \frac{\d}{\d t} \hat{\rho}_S(t) &= -i \big[ \hat{H}_S, \hat{\rho}_S(t) \big] - \frac{1}{2i\lambda} \sum_{\alpha \in \mathcal{I}} \bigg( \Big[ \big( \hat{U}_\lambda - \hat{U}_\lambda^\dagger \big), \, \hat{B}_\alpha(t,t_0) \hat{\rho}_S(t) \Big] + \Big[ \hat{\rho}_S(t) \hat{C}_\alpha(t,t_0), \, \big( \hat{U}_\lambda - \hat{U}_\lambda^\dagger \big) \Big] \bigg).
\end{align*}

This is one of the standard forms of the equation, however when it comes to attributing physical meaning to the equation, it is more feasible to cast it into a different form, expressed as a combination of double-commutators and commutators of anti-commutators, respectively. For this purpose, we use the following identities:

\begin{align*}
    \Big[ \big( \hat{U}_\lambda - \hat{U}_\lambda^\dagger \big), \, \big( \hat{U}_\lambda(-\xi) - \hat{U}_\lambda^\dagger(-\xi) \big) \hat{\rho}_S(t) \Big] = \frac{1}{2} \bigg( &\Big[ \big( \hat{U}_\lambda - \hat{U}_\lambda^\dagger \big), \Big[ \big( \hat{U}_\lambda(-\xi) - \hat{U}_\lambda^\dagger(-\xi) \big), \hat{\rho}_S(t) \Big] \Big] \\
    +&\Big[ \big( \hat{U}_\lambda - \hat{U}_\lambda^\dagger \big), \Big\{ \big( \hat{U}_\lambda(-\xi) - \hat{U}_\lambda^\dagger(-\xi) \big), \hat{\rho}_S(t) \Big\} \Big] \bigg), \\[0.7em]
    \Big[ \hat{\rho}_S(t) \big( \hat{U}_\lambda(-\xi) - \hat{U}_\lambda^\dagger(-\xi) \big), \, \big( \hat{U}_\lambda - \hat{U}_\lambda^\dagger \big) \Big] = \frac{1}{2} \bigg( &\Big[ \big( \hat{U}_\lambda - \hat{U}_\lambda^\dagger \big), \Big[ \big( \hat{U}_\lambda(-\xi) - \hat{U}_\lambda^\dagger(-\xi) \big), \hat{\rho}_S(t) \Big] \Big] \\
    -&\Big[ \big( \hat{U}_\lambda - \hat{U}_\lambda^\dagger \big), \Big\{ \big( \hat{U}_\lambda(-\xi) - \hat{U}_\lambda^\dagger(-\xi) \big), \hat{\rho}_S(t) \Big\} \Big] \bigg),
\end{align*}

where $\{\cdot,\cdot\}$ denotes the anticommutator. Note that the above two expressions only differ by the sign in front of the commutator-anticommutator combination, which precisely fits the difference in the sign of the argument of $C_{\alpha \alpha}$ in the definitions of the operators $\hat{B}_\alpha(t,t_0)$ and $\hat{C}_\alpha(t,t_0)$ in (\ref{eq:B_Coeff}) and (\ref{eq:C_Coeff}), respectively. This lets us immediately conclude that in the final master equation, the double-commutator is smeared with the cosine contribution whereas the commutator-anticommutator term is smeared with the sine-contribution from the environmental correlation functions, the mixed contributions cancel essentially by virtue of the way the correlation functions are defined. If we now gather all terms, we get the following expression for the final master equation of the open $U(1)$ scattering model:

\begin{align}
    \frac{\d}{\d t} \hat{\rho}_S(t) = -i [\hat{H}_S, \hat{\rho}_S(t)] &+ \frac{1}{8 \lambda^4} \int\limits_{0}^{t-t_0} \d \xi \, \sum_{\alpha=1}^N \, \sum_{n \in \mathbb{Z}} g_\alpha^2 \frac{e^{-\sigma_\alpha n^2} e^{-2 \sigma_\alpha n \delta}}{\vartheta(z_\alpha, \tau_\alpha)} \cos \Big(\frac{\xi}{m_\alpha} \lambda(n+\delta) \Big) \label{eq:U1_master}\\
    &\times \bigg( \cos\Big( \frac{\xi \lambda^2}{2 m_\alpha} \Big) \Big[ \hat{U}_\lambda - \hat{U}_\lambda^\dagger , \Big[ \hat{U}_\lambda(-\xi) - \hat{U}_\lambda^\dagger(-\xi), \hat{\rho}_S(t) \Big] \Big] \nonumber \\
    -&\;i\sin \Big( \frac{\xi \lambda^2}{2 m_\alpha} \Big) \Big[ \hat{U}_\lambda - \hat{U}_\lambda^\dagger , \Big\{ \hat{U}_\lambda(-\xi) - \hat{U}_\lambda^\dagger(-\xi), \hat{\rho}_S(t) \Big\} \Big] \bigg). \nonumber
\end{align}

In the limit of a vanishing 'temperature' parameter $T$ we reduce the initial environmental density operator to a pure state and correspondingly are left with the vacuum contribution $n=0$ instead of the sum over all integers:

\begin{align*}
    \lim_{T \to 0} \Big( \frac{\d}{\d t} \hat{\rho}_S(t) \Big) = -i [\hat{H}_S, \hat{\rho}_S(t)] &+ \frac{1}{8 \lambda^4} \int\limits_{0}^{t-t_0} \d \xi \, \sum_{\alpha=1}^N g_\alpha^2 \cos\Big( \frac{\xi}{m_\alpha} \lambda\delta \Big) \times \\
    &\bigg( \cos\Big( \frac{\xi \lambda^2}{2 m_\alpha} \Big) \Big[ \hat{U}_\lambda - \hat{U}_\lambda^\dagger , \Big[ \hat{U}_\lambda(-\xi) - \hat{U}_\lambda^\dagger(-\xi), \hat{\rho}_S(t) \Big] \Big] \\
    -&\;i\sin \Big( \frac{\xi \lambda^2}{2 m_\alpha} \Big) \Big[ \hat{U}_\lambda - \hat{U}_\lambda^\dagger , \Big\{ \hat{U}_\lambda(-\xi) - \hat{U}_\lambda^\dagger(-\xi), \hat{\rho}_S(t) \Big\} \Big] \bigg).
\end{align*}

\subsection{Physical Properties of the Master Equation of the Open $U(1)$ Scattering Model} \label{sec:model_interpretation}

In the context of the four canonical models of open quantum systems \cite{Breuer:OpenSystems}, there is a specific interpretation to the individual terms in the master equation, see for instance~\cite{Caldeira:QBM}. The~nomenclature of the real and imaginary part of the environmental correlation functions as noise- and dissipation kernel already are suggestive. It is indeed the noise kernel that usually determines the decoherence dynamics. Suppose we are interested in the temporal derivative of the expectation value of the system's momentum. Then we can replace the expectation value with a trace and apply the derivative to the density operator and consequently replace this derivative by the right hand side of the master equation. The~resulting equation can be greatly simplified by realizing that the trace is cyclic. This is true in our case and generally needs to be checked explicitly if unbounded operators are involved. Under the assumptions of the validity of the Born--Markov master equation one can deduce a system of generally coupled differential equations for the various moments of the position and momentum variables, respectively. Let us quickly recall the case of collisional decoherence as discussed in the introduction and the form of its master equation derived in \cite{hornberger2003collisional,busse2009pointer}:

\begin{equation}
    \frac{\d}{\d t} \hat{\rho}(t) = -i [\hat{H}, \hat{\rho}(t)] + \int\limits_{\mathbb{R}} \d a \; \Big(\hat{L}_a \hat{\rho}(t) \hat{L}_a^\dagger - \frac{1}{2} \hat{L}_a \hat{L}_a^\dagger \hat{\rho}(t) - \frac{1}{2} \hat{\rho}(t) \hat{L}_a \hat{L}_a^\dagger \Big), \quad \hat{L}_a \defeq \sqrt{\gamma G(a)} e^{ia\hat{q}}, \label{eq:busse_collisional_dec}
\end{equation}

where $G(a) \geq 0$ is the momentum transfer function with $\int \d a \; G(a) = 1$ and some collision rate $\gamma$. It is immediately clear that this equation has Lindblad form, which is evidently not the case for the equation derived here (\ref{eq:U1_master}), since there is a residual time-dependence in the effective system operators. In case of $\langle \hat{q} \rangle_\rho$ and $\langle \hat{p} \rangle_\rho$, where the expectation values with respect to the system's density operator has been denoted with a lowercase $_\rho$ for more clarity, the resulting differential equations in the collisional decoherence model amount to: 

\begin{equation}
    \frac{\d}{\d t}\langle \hat{q} \rangle_\rho = \frac{\langle \hat{p} \rangle_\rho}{m}, \qquad \frac{\d}{\d t}\langle \hat{p} \rangle_\rho = \gamma \int\limits_\mathbb{R} \d a \; a G(a), \label{eq:busse_int}
\end{equation}
which vanishes in the case of a symmetric function $G(a) = G(-a)$, suggesting $\langle \hat{p} \rangle_\rho \sim \textrm{const}$. A similar computation \cite{hornberger2003collisional,busse2009pointer} for the position operator yields $m \, \partial_t \langle \hat{q} \rangle_\rho = \langle \hat{p} \rangle_\rho$, just as one would expect based on intuition from classical physics. We now shall repeat this analysis for the master equation in the previous section and derive the equations of motion for the respective first moments. For now it is sufficient to express the expectation value in terms of proportionalities, at the end of the derivation we will again collect all proper prefactors. We obtain

\begin{align*}
    \frac{\d}{\d t} \big\langle \hat{p}_\varphi \big\rangle_\rho = \Tr(\hat{p}_\varphi  \frac{\d}{\d t} \hat{\rho}_S(t)) &\sim \cos\Big( \frac{\xi \lambda^2}{2 m_\alpha} \Big) \Tr( \hat{p}_\varphi \Big[ \hat{U}_\lambda - \hat{U}_\lambda^\dagger , \Big[ \hat{U}_\lambda(-\xi) - \hat{U}_\lambda^\dagger(-\xi), \hat{\rho}_S(t) \Big] \Big]) \\
    &- i\sin \Big( \frac{\xi \lambda^2}{2 m_\alpha} \Big)\Tr(\hat{p}_\varphi  \Big[ \hat{U}_\lambda - \hat{U}_\lambda^\dagger , \Big\{ \hat{U}_\lambda(-\xi) - \hat{U}_\lambda^\dagger(-\xi), \hat{\rho}_S(t) \Big\} \Big]).
\end{align*}

Hence, we are interested in the explicit expressions for $\Tr( \hat{p}_\varphi \big[\hat{Q}_\lambda, \big[ \hat{Q}_\lambda(-\xi), \hat{\rho}_S(t) \big] \big])$ and $\Tr{ \hat{p}_\varphi \big[\hat{Q}_\lambda, \big\{ \hat{Q}_\lambda(-\xi), \hat{\rho}_S(t) \big\} \big]}$ respectively, where we set $\hat{U}_\lambda - \hat{U}_\lambda^\dagger \eqdef 2i\lambda \hat{Q}_\lambda$ for brevity. Note that the unitary evolution generated by the free particle Hamiltonian $\hat{H}_S$ does not contribute in this computation since $\hat{p}_\varphi$ commutes with itself and the trace is invariant under cyclic permutations of its arguments. This yields

\begin{align*}
    \Tr( \hat{p}_\varphi \big[\hat{Q}_\lambda, \big[ \hat{Q}_\lambda(-\xi), \hat{\rho}_S \big] \big]) 
    &= \Tr(\Big[ \hat{Q}_\lambda(-\xi) \big[\hat{Q}_\lambda, \hat{p}_\varphi \big] + \big[ \hat{p}_\varphi, \hat{Q}_\lambda \big] \hat{Q}_\lambda(-\xi) \Big] \hat{\rho}_S ) \\[0.3em]
    &= \frac{1}{2i} \Tr(\Big[ \Big( \hat{U}_\lambda + \hat{U}_\lambda^\dagger \Big) \hat{Q}_\lambda(-\xi) - \hat{Q}_\lambda(-\xi) \Big( \hat{U}_\lambda + \hat{U}_\lambda^\dagger \Big) \Big] \hat{\rho}_S ) \\[0.3em]
    &= \frac{1}{2i} \Tr( \Big[ \hat{U}_\lambda + \hat{U}_\lambda^\dagger, \hat{Q}_\lambda(-\xi) \Big] \hat{\rho}_S ),
\end{align*}
where we have first permuted every instance of $\hat{\rho}_S$ to the rightmost position and consecutively cancelled the first and fourth term in the trace by inserting the commutators. As a last step we used $2i \big[ \hat{p}_\varphi, \hat{Q}_\lambda \big] = (\hat{U}_\lambda + \hat{U}_\lambda^\dagger)$. On a similar note we can simplify the second contribution:

\begin{align*}
    \Tr( \hat{p}_\varphi \big[\hat{Q}_\lambda, \big\{ \hat{Q}_\lambda(-\xi), \hat{\rho}_S \big\} \big]) 
    &= \Tr(\Big[ \big[ \hat{p}_\varphi, \hat{Q}_\lambda \big] \hat{Q}_\lambda(-\xi) - \hat{Q}_\lambda(-\xi) \big[\hat{Q}_\lambda, \hat{p}_\varphi \big] \Big] \hat{\rho}_S ) \\[0.3em]
    &= \frac{1}{2i} \Tr( \Big[ \Big( \hat{U}_\lambda + \hat{U}_\lambda^\dagger \Big) \hat{Q}_\lambda(-\xi) + \hat{Q}_\lambda(-\xi) \Big( \hat{U}_\lambda + \hat{U}_\lambda^\dagger \Big) \Big] \hat{\rho}_S ) \\[0.3em]
    &= \frac{1}{2i} \Tr( \Big\{ \hat{U}_\lambda + \hat{U}_\lambda^\dagger, \hat{Q}_\lambda(-\xi) \Big\} \hat{\rho}_S ).
\end{align*}

Note that we left the interaction picture's operators in $\hat{Q}_\lambda(-\xi)$ untouched so far, the remaining trace can now be evaluated with the Weyl relations between the one-parameter unitary groups associated to the canonical variables known from ordinary quantum mechanics. More precisely, we need to simplify commutators of four different kinds:

\begin{align*}
    \Big[ \hat{U}_\lambda, \hat{U}_\lambda(-\xi) \Big] &= e^{-\frac{i \xi \lambda^2}{2 m_0}} \hat{U}_\lambda \Big[ \hat{U}_\lambda, \exp\Big\{-\frac{i \xi \lambda}{m_0} \hat{p}_\varphi \Big\} \Big] = e^{-\frac{i \xi \lambda^2}{2 m_0}} \Big( 1 - e^{ -\frac{i \xi \lambda^2}{m_0}} \Big) \hat{U}_\lambda^2 \exp\Big\{- \frac{i \xi \lambda}{m_0} \hat{p}_\varphi \Big\}, \\[0.3em]
    \Big[ \hat{U}_\lambda^\dagger, \hat{U}_\lambda(-\xi) \Big] &= e^{-\frac{i \xi \lambda^2}{2 m_0}} \hat{U}_\lambda \Big[ \hat{U}_\lambda^\dagger, \exp\Big\{-\frac{i \xi \lambda}{m_0} \hat{p}_\varphi \Big\} \Big] = e^{-\frac{i \xi \lambda^2}{2 m_0}} \Big( 1 - e^{\frac{i \xi \lambda^2}{m_0}} \Big) \hat{U}_\lambda \hat{U}_\lambda^\dagger \exp\Big\{-\frac{i \xi \lambda}{m_0} \hat{p}_\varphi \Big\}, \\[0.3em]
    \Big[ \hat{U}_\lambda, \hat{U}_\lambda^\dagger(-\xi) \Big] &= e^{-\frac{i \xi \lambda^2}{2 m_0}} \hat{U}_\lambda^\dagger \Big[ \hat{U}_\lambda, \exp\Big\{\frac{i \xi \lambda}{m_0} \hat{p}_\varphi \Big\} \Big] = e^{-\frac{i \xi \lambda^2}{2 m_0}} \Big( 1 - e^{\frac{i \xi \lambda^2}{m_0}} \Big) \hat{U}_\lambda^\dagger \hat{U}_\lambda \exp\Big\{\frac{i \xi \lambda}{m_0} \hat{p}_\varphi \Big\}, \\[0.3em]
    \Big[ \hat{U}_\lambda^\dagger, \hat{U}_\lambda^\dagger(-\xi) \Big] &= e^{-\frac{i \xi \lambda^2}{2 m_0}} \hat{U}_\lambda^\dagger \Big[ \hat{U}_\lambda^\dagger, \exp\Big\{\frac{i \xi \lambda}{m_0} \hat{p}_\varphi \Big\} \Big] = e^{-\frac{i \xi \lambda^2}{2 m_0}} \Big( 1 - e^{-\frac{i \xi \lambda^2}{m_0}} \Big)  \big( \hat{U}_\lambda^\dagger \big)^2 \exp\Big\{\frac{i \xi \lambda}{m_0} \hat{p}_\varphi \Big\},
\end{align*}

where we repeatedly used the Weyl relations to rewrite the commutators in the second step of each line. The~exponential prefactor is universal to \textit{both} operators $\hat{U}_\lambda(-\xi)$ and $\hat{U}_\lambda^\dagger(-\xi)$, which is evident if we recall the operator ordering that we chose beforehand and the fact that these operators contain an exponential of $\hat{p}_\varphi$ with which neither $\hat{U}_\lambda$ nor $\hat{U}_\lambda^\dagger$ commute. On a similar note the anticommutators can be obtained by carefully rearranging terms, using the Weyl relations and the fact that $\hat{U}_\lambda$ and $\hat{U}_\lambda^\dagger$ commute among themselves

\begin{align*}
    \Big\{ \hat{U}_\lambda, \hat{U}_\lambda(-\xi) \Big\} &= e^{-\frac{i \xi \lambda^2}{2 m_0}} \Big( 1 + e^{ -\frac{i \xi \lambda^2}{m_0}} \Big) \hat{U}_\lambda^2 \exp\Big\{- \frac{i \xi \lambda}{m_0} \hat{p}_\varphi \Big\}, \\[0.3em]
    \Big\{ \hat{U}_\lambda^\dagger, \hat{U}_\lambda(-\xi) \Big\} &= e^{-\frac{i \xi \lambda^2}{2 m_0}} \Big( 1 + e^{\frac{i \xi \lambda^2}{m_0}} \Big) \hat{U}_\lambda \hat{U}_\lambda^\dagger \exp\Big\{-\frac{i \xi \lambda}{m_0} \hat{p}_\varphi \Big\}, \\[0.3em]
    \Big\{ \hat{U}_\lambda, \hat{U}_\lambda^\dagger(-\xi) \Big\} &= e^{-\frac{i \xi \lambda^2}{2 m_0}} \Big( 1 + e^{\frac{i \xi \lambda^2}{m_0}} \Big) \hat{U}_\lambda^\dagger \hat{U}_\lambda \exp\Big\{\frac{i \xi \lambda}{m_0} \hat{p}_\varphi \Big\}, \\[0.3em]
    \Big\{ \hat{U}_\lambda^\dagger, \hat{U}_\lambda^\dagger(-\xi) \Big\} &= e^{-\frac{i \xi \lambda^2}{2 m_0}} \Big( 1 + e^{-\frac{i \xi \lambda^2}{m_0}} \Big)  \big( \hat{U}_\lambda^\dagger \big)^2 \exp\Big\{\frac{i \xi \lambda}{m_0} \hat{p}_\varphi \Big\}.
\end{align*}

If we collect all contributions for the commutator we end up with

\begin{align*} \hspace{-0.3cm}
     2i\lambda \Tr(\Big[ \hat{U}_\lambda + \hat{U}_\lambda^\dagger, \hat{Q}_\lambda(-\xi) \Big] \hat{\rho}_S ) &= e^{-\frac{i \xi \lambda^2}{2 m_0}} \Big(1-e^{\frac{i \xi \lambda^2}{m_0}} \Big) \Big\langle \exp\Big\{\frac{-i \xi \lambda}{m_0} \hat{p}_\varphi \Big\} - \exp\Big\{\frac{i \xi \lambda}{m_0} \hat{p}_\varphi \Big\} \Big\rangle_\rho \\
     &+ e^{-\frac{i \xi \lambda^2}{2 m_0}} \Big(1-e^{\frac{-i \xi \lambda^2}{m_0}} \Big) \Big\langle \hat{U}_\lambda^2 \exp\Big\{\frac{-i \xi \lambda}{m_0} \hat{p}_\varphi \Big\} - \big( \hat{U}_\lambda^\dagger \big)^2 \exp\Big\{\frac{i \xi \lambda}{m_0} \hat{p}_\varphi \Big\} \Big\rangle_\rho.
\end{align*}

and in the case of the contribution involving the anticommutator we get

\begin{align*} \hspace{-0.3cm}
    2i \lambda \Tr( \Big\{ \hat{U}_\lambda + \hat{U}_\lambda^\dagger, \hat{Q}_\lambda(-\xi) \Big\} \hat{\rho}_S ) &= e^{-\frac{i \xi \lambda^2}{2 m_0}} \Big(1+e^{\frac{i \xi \lambda^2}{m_0}} \Big) \Big\langle \exp\Big\{\frac{-i \xi \lambda}{m_0} \hat{p}_\varphi \Big\} - \exp\Big\{\frac{i \xi \lambda}{m_0} \hat{p}_\varphi \Big\} \Big\rangle_\rho \\
     &+ e^{-\frac{i \xi \lambda^2}{2 m_0}} \Big(1+e^{\frac{-i \xi \lambda^2}{m_0}} \Big) \Big\langle \hat{U}_\lambda^2 \exp\Big\{\frac{-i \xi \lambda}{m_0} \hat{p}_\varphi \Big\} - \big( \hat{U}_\lambda^\dagger \big)^2 \exp\Big\{\frac{i \xi \lambda}{m_0} \hat{p}_\varphi \Big\} \Big\rangle_\rho.
\end{align*}

On top of this, the expectation value of the commutator and anticommutator brackets that do not include the holonomies can be further simplified to represent an analytic function of $\langle \hat{p}_\varphi \rangle_\rho$ when we consider $U(1)$ complexifier coherent states \cite{Bahr:Coherent,thiemann2006complexifier}. In the course of our analysis, we do not observe the higher-order corrections from a more general class of states due to the truncation at linear order in $\lambda$ performed in consecutive steps, hence we opt for the more compact notation at this point. Furthermore, let us abbreviate the second expectation value with $\langle \hat{\Lambda} \rangle_\rho$, that is:

\begin{align*}
    \Big\langle \exp\Big\{\frac{-i \xi \lambda}{m_0} \hat{p}_\varphi \Big\} &- \exp\Big\{\frac{i \xi \lambda}{m_0} \hat{p}_\varphi \Big\} \Big\rangle_\rho = -2i \Big\langle \sin\Big(\frac{i \lambda \xi}{m_0} \hat{p}_\varphi \Big) \Big\rangle_\rho = -2i \sin\Big(\frac{i \lambda \xi}{m_0} \big\langle \hat{p}_\varphi \big\rangle_\rho \Big), \\[0.4em]
    &\Big\langle \hat{U}_\lambda^2 \exp\Big\{\frac{-i \xi \lambda}{m_0} \hat{p}_\varphi \Big\} - \big( \hat{U}_\lambda^\dagger \big)^2 \exp\Big\{\frac{i \xi \lambda}{m_0} \hat{p}_\varphi \Big\} \Big\rangle_\rho \defeq \langle \hat{\Lambda} \rangle_\rho.
\end{align*}

Hence, the effective equation of motion for the expectation value of $\hat{p}_\varphi$ is given by:

\begin{align}
    &\frac{\d}{\d t} \big\langle \hat{p}_\varphi \big\rangle_\rho = \frac{1}{8 \lambda^4} \int\limits_{0}^{t-t_0} \d \xi \, \sum_{\alpha=1}^N \, \sum_{n \in \mathbb{Z}} g_\alpha^2 \frac{e^{-\sigma_\alpha n^2} e^{-2 \sigma_\alpha n \delta}}{\vartheta(z_\alpha, \tau_\alpha)} \cos \Big(\frac{\xi}{m_\alpha} \lambda(n+\delta) \Big) \times \nonumber \\
    \bigg[& (-2i)^2 \lambda \cos\Big( \frac{\xi \lambda^2}{2 m_\alpha}\Big) \sin\Big( \frac{\xi \lambda^2}{2 m_0}\Big) \Big\langle \exp\Big\{\frac{i \lambda \xi}{m_0} \hat{p}_\varphi\Big\} - \exp\Big\{ -\frac{i \lambda \xi}{m_0} \hat{p}_\varphi\Big\} \Big\rangle_\rho \nonumber \\
    +&(-2i)^2 \lambda \sin \Big( \frac{\xi \lambda^2}{2 m_\alpha} \Big) \cos\Big( \frac{\xi \lambda^2}{2 m_0}\Big) \Big\langle \exp\Big\{\frac{i \lambda \xi}{m_0} \hat{p}_\varphi\Big\} - \exp\Big\{ -\frac{i \lambda \xi}{m_0} \hat{p}_\varphi\Big\} \Big\rangle_\rho \bigg) \nonumber \\
    +&2 \lambda \cos\Big( \frac{\xi \lambda^2}{2 m_\alpha}\Big) \sin\Big( \frac{\xi \lambda^2}{2 m_0}\Big) \bigg( \sin\Big( \frac{\xi \lambda^2}{m_0}\Big) + i \cos\Big( \frac{\xi \lambda^2}{m_0}\Big) \bigg) \langle \hat{\Lambda} \rangle_\rho \nonumber \\
    -& 2i \lambda \sin\Big( \frac{\xi \lambda^2}{2 m_\alpha}\Big) \cos\Big( \frac{\xi \lambda^2}{2 m_0}\Big) \bigg( \cos\Big( \frac{\xi \lambda^2}{m_0}\Big) - i \sin\Big( \frac{\xi \lambda^2}{m_0}\Big) \bigg) \langle \hat{\Lambda} \rangle_\rho \bigg] \nonumber \\
    &= - \frac{1}{2 \lambda^3} \int\limits_{0}^{t-t_0} \d \xi \, \sum_{\alpha=1}^N \, \sum_{n \in \mathbb{Z}} g_\alpha^2 \frac{e^{-\sigma_\alpha n^2} e^{-2 \sigma_\alpha n \delta}}{\vartheta(z_\alpha, \tau_\alpha)} \cos \Big(\frac{\xi}{m_\alpha} \lambda(n+\delta) \Big) \times \label{eq:gen_expval_p}\\
    \bigg[& \bigg( \cos\Big( \frac{\xi \lambda^2}{2 m_\alpha}\Big) \sin\Big( \frac{\xi \lambda^2}{2 m_0}\Big) + \sin \Big( \frac{\xi \lambda^2}{2 m_\alpha} \Big) \cos\Big( \frac{\xi \lambda^2}{2 m_0}\Big) \bigg) \sin\Big(\frac{\lambda \xi}{m_0} \big\langle \hat{p}_\varphi \big\rangle_\rho \Big) \nonumber \\
    -&\frac{1}{2} \cos\Big( \frac{\xi \lambda^2}{2 m_\alpha}\Big) \sin\Big( \frac{\xi \lambda^2}{2 m_0}\Big) \bigg( \sin\Big( \frac{\xi \lambda^2}{m_0}\Big) + i \cos\Big( \frac{\xi \lambda^2}{m_0}\Big) \bigg) \langle \hat{\Lambda} \rangle_\rho \nonumber \\
    +& \frac{i}{2}\sin\Big( \frac{\xi \lambda^2}{2 m_\alpha}\Big) \cos\Big( \frac{\xi \lambda^2}{2 m_0}\Big) \bigg( \cos\Big( \frac{\xi \lambda^2}{m_0}\Big) - i \sin\Big( \frac{\xi \lambda^2}{m_0}\Big) \bigg) \langle \hat{\Lambda} \rangle_\rho \bigg]. \nonumber
\end{align}

In order to get a better understanding of the properties of this equation we would like to expand the prefactors of the expectation values in terms of $\lambda$. This will establish a correspondence between the momentum transfer function $G(a)$ in \cite{busse2009pointer} and the choice of our environmental interaction Hamiltonian. We treat each line individually, starting with the first one:

\begin{align*}
   &\frac{1}{2 \lambda^3} \cos \Big(\frac{\xi}{m_\alpha} \lambda(n+\delta) \Big) \bigg( \cos\Big( \frac{\xi \lambda^2}{2 m_\alpha}\Big) \sin\Big( \frac{\xi \lambda^2}{2 m_0}\Big) + \sin \Big( \frac{\xi \lambda^2}{2 m_\alpha} \Big) \cos\Big( \frac{\xi \lambda^2}{2 m_0}\Big) \bigg) \sin\Big(\frac{\lambda\xi}{m_0} \big\langle \hat{p}_\varphi \big\rangle_\rho \Big) \\
   = &\frac{1}{2 \lambda^3} \bigg[ \bigg( \Big( \frac{\xi \lambda^2}{2 m_0}\Big) + \Big( \frac{\xi \lambda^2}{2 m_\alpha}\Big) \bigg) \frac{\lambda\xi}{m_0} \big\langle \hat{p}_\varphi \big\rangle_\rho + \mathcal{O}(\lambda^5) \bigg]
   = \frac{\xi^2}{4 m_0} \bigg( \frac{1}{m_0} + \frac{1}{m_\alpha} \bigg)\big\langle \hat{p}_\varphi \big\rangle_\rho + \mathcal{O}(\lambda).
\end{align*}

There is no contribution of lower (or even inverse) order since $\sin(\lambda) \sim \lambda + \mathcal{O}(\lambda^3)$ and both summands in the bracket contain such a contribution with $\lambda^2$ in the argument. The~superselection parameter $\delta$ does start to contribute in $\mathcal{O}(\lambda^2)$ since $\cos(\lambda) \sim \lambda^2 + \mathcal{O}(\lambda^4)$ and the bracket together with the expectation value is at least of order $\mathcal{O}(\lambda)$. We would like to stress that this part of the master equation was particularly simple to evaluate since there was no involvement of the holonomies whatsoever. An expansion of $\hat{U}_\lambda$ and $\hat{U}_\lambda^\dagger$ is not meaningful in the quantum framework since the associated generator does not exist. The~strategy will now be to only expand the prefactor of the expectation values including holonomies and investigate the relation between the noise- and dissipation kernel order by order in the Taylor series. We get

\begin{align*}
    \frac{1}{4 \lambda^3} \cos\Big( \frac{\xi \lambda^2}{2 m_\alpha}\Big) \sin\Big( \frac{\xi \lambda^2}{2 m_0}\Big) \bigg( \sin\Big( \frac{\xi \lambda^2}{m_0}\Big) + i \cos\Big( \frac{\xi \lambda^2}{m_0}\Big) \bigg) \big\langle \hat{\Lambda} \big\rangle_\rho = \frac{i}{8} \frac{\xi}{m_0 \lambda} \big\langle \hat{\Lambda} \big\rangle_\rho + \mathcal{O}(\lambda),\\[0.4em]
    \frac{i}{4 \lambda^3} \sin\Big( \frac{\xi \lambda^2}{2 m_\alpha}\Big) \cos\Big( \frac{\xi \lambda^2}{2 m_0}\Big) \bigg( \cos\Big( \frac{\xi \lambda^2}{m_0}\Big) - i \sin\Big( \frac{\xi \lambda^2}{m_0}\Big) \bigg) \big\langle \hat{\Lambda} \big\rangle_\rho = \frac{i}{8} \frac{\xi}{m_\alpha \lambda} \big\langle \hat{\Lambda} \big\rangle_\rho + \mathcal{O}(\lambda),
\end{align*}

Note that these expansions only differ in the mass label $m_\alpha$ and hence generally do not cancel.  It is not possible to take the limit $\lambda \to 0$ in the chosen representation for any $m_\alpha \neq m_0$ as the expansion of $\big\langle \hat{\Lambda} \big\rangle_\rho$ will contain a zeroth order contribution proportional (but not limited) to $\sim \big\langle 4i \varphi \big\rangle_\rho + \mathcal{O}(\lambda)$, which are at best meaningful in a semiclassical regime because of the discontinuity of the holonomy operators in the superselection sectors due to  which the angle operator cannot be implemented on our Hilbert space. This is the point where our result can be related to the symmetry properties of the momentum transfer function found in \cite{busse2009pointer}, the case of a symmetric momentum transfer function in~(\ref{eq:busse_int}) and hence $\big\langle \hat{p}_\varphi \big\rangle_\rho = \textrm{const.}$ is \textit{not} included in the polymerized scattering model for a non-vanishing coupling constant. Even the lowest-order contribution of $\big\langle \hat{p}_\varphi \big\rangle_\rho$ shows a different behaviour than Schrödinger-type models. Henceforth, we assume that the environment consists of particles with masses identical to the system mass $m_0$, which lets us directly simplify the effective differential equation for $\big\langle \hat{p}_\varphi \big\rangle_\rho$ since all problematic terms cancel. The conventional approach suggests that the sum over couplings and masses is replaced by an integral over a so-called a-priori \textit{spectral} \textit{density} \cite{Breuer:OpenSystems}. This in a sense corresponds to the thermodynamical limit, the environmental modes (or masses, respectively) are assumed to lie dense enough as to be replaced by a continuous integral instead of a sum. The~\textit{choice} of spectral density then crucially influences the properties of the model \cite{Schlosshauer:Boson-Spin} but can be used to remedy the occurrence of spontaneous \textit{re}coherence or \textit{Poincaré recurrences}, respectively) cancel up to arbitrary order in $\lambda$:

\begin{align}
    \frac{\d}{\d t} \big\langle \hat{p}_\varphi \big\rangle_\rho &=  - \frac{1}{\lambda^3} \int\limits_{0}^{t-t_0} \d \xi \, \sum_{\stackrel{\alpha=1}{n \in \mathbb{Z}}}^N g_\alpha^2 \frac{e^{-\sigma n^2} e^{-2 \sigma n \delta}}{\vartheta(z, \tau)} \cos \Big(\frac{\xi}{m_0} \lambda(n+\delta) \Big)
    \cos\Big( \frac{\xi \lambda^2}{2 m_0}\Big) \sin\Big( \frac{\xi \lambda^2}{2 m_0}\Big) \sin\Big(\frac{\lambda \xi}{m_0} \big\langle \hat{p}_\varphi \big\rangle_\rho \Big) \nonumber \\
    &= - \int\limits_{0}^{t-t_0} \d \xi \, \sum_{\alpha=1}^N \, \sum_{n \in \mathbb{Z}} g_\alpha^2 \frac{e^{-\sigma n^2} e^{-2 \sigma n \delta}}{\vartheta(z, \tau)} \frac{\xi^2}{2 m_0^2}\big\langle \hat{p}_\varphi \big\rangle_\rho + \mathcal{O}(\lambda^2) \nonumber \\
    &= - \sum_{\alpha=1}^N \, \sum_{n \in \mathbb{Z}} g_\alpha^2 \frac{e^{-\sigma n^2} e^{-2 \sigma n \delta}}{\vartheta(z, \tau)} \frac{(t-t_0)^3}{6 m_0^2} \big\langle \hat{p}_\varphi \big\rangle_\rho + \mathcal{O}(\lambda^2) \nonumber \\
    &= - G \frac{(t-t_0)^3}{6 m_0^2} \big\langle \hat{p}_\varphi \big\rangle_\rho + \mathcal{O}(\lambda^2), \quad \textrm{with} \quad G \defeq \sum_{\alpha=1}^N g_\alpha^2 \geq 0 \quad \forall \quad N < \infty.
    \label{eq:expvalue_p}
\end{align}

This can immediately be integrated to yield the lowest-order solution for $\big\langle \hat{p}_\varphi \big\rangle_\rho$:

\begin{equation}
    \big\langle \hat{p}_\varphi \big\rangle_\rho(t) = \exp\bigg\{ - G \frac{(t-t_0)^4}{24 m_0^2} \bigg\} \big\langle \hat{p}_\varphi \big\rangle_\rho(t_0). \label{eq:solution_p}
\end{equation}

Hence, we get a strong suppression of the momentum expectation value even in the lowest-order expansion in $\lambda$. The~case of $\langle \hat{Q}_\lambda \rangle_\rho$ is significantly easier. We can immediately see that the contribution from the non-unitary term in the master equation vanishes by virtue of the cyclicity of the trace, which is evident from the boundedness of every operator~involved:

\begin{align*}
    \Tr( \hat{Q}_\lambda \big[\hat{Q}_\lambda, \big[ \hat{Q}_\lambda(-\xi), \hat{\rho}_S \big] \big]) 
    &= \Tr( \Big( \hat{Q}_\lambda^2 \hat{Q}_\lambda(-\xi) - \hat{Q}_\lambda(-\xi) \hat{Q}_\lambda^2 - \hat{Q}_\lambda^2 \hat{Q}_\lambda(-\xi) + \hat{Q}_\lambda(-\xi) \hat{Q}_\lambda^2 \Big) \hat{\rho}_S) = 0.
\end{align*}

A similar argument holds for the anticommutator:

\begin{align*}
    \Tr( \hat{Q}_\lambda \big[\hat{Q}_\lambda, \big\{ \hat{Q}_\lambda(-\xi), \hat{\rho}_S \big\} \big]) 
    &= \Tr( \Big( \hat{Q}_\lambda^2 \hat{Q}_\lambda(-\xi) + \hat{Q}_\lambda(-\xi) \hat{Q}_\lambda^2 - \hat{Q}_\lambda^2 \hat{Q}_\lambda(-\xi) - \hat{Q}_\lambda(-\xi) \hat{Q}_\lambda^2 \Big) \hat{\rho}_S) = 0.
\end{align*}

The~only non-vanishing term comes from the unitary evolution:

\begin{align}
    \frac{\d}{\d t} \big\langle \hat{Q}_\lambda \big\rangle_\rho &= -i \Tr( \hat{Q}_\lambda \big[ \hat{H}_S, \hat{\rho}_S(t) \big]) = -\frac{i}{2 m_0} \Tr( \big[\hat{Q}_\lambda, \hat{p}_\varphi^2\big] \hat{\rho}_S) \label{eq:eff_Q}\\
    &= -\frac{1}{4 \lambda m_0} \big\langle \big[\hat{U}_\lambda, \hat{p}_\varphi^2 \big] - \big[\hat{U}_\lambda^\dagger, \hat{p}_\varphi^2 \big]\big\rangle_\rho \nonumber \\
    &= \frac{1}{4 m_0} \big\langle \big(\hat{U}_\lambda + \hat{U}_\lambda^\dagger \big) \hat{p}_\varphi + \hat{p}_\varphi \big(\hat{U}_\lambda + \hat{U}_\lambda^\dagger \big) \big\rangle_\rho \nonumber \\
    &= \frac{1}{2 m_0} \big\langle \widehat{\cos(\lambda\varphi)} \hat{p}_\varphi + \hat{p}_\varphi \widehat{\cos(\lambda\varphi)} \big\rangle_\rho.
\end{align}

As can be seen the resulting equation can not be solely written in terms of expectation values of $\hat{p}_\varphi$ and $\hat{Q}_\lambda$ due to the non-closing of the algebra between these two variables, since the relevant algebra is formed by $\big(\hat{p}_\varphi, \hat{U}_\lambda, \hat{U}_\lambda^\dagger \big)$ and not combinations of the latter two. We can however attempt to solve Equation (\ref{eq:eff_Q}) for $\big\langle \hat{Q}_\lambda \big\rangle_\rho$ based on the lowest-order solution of $\big\langle \hat{p}_\varphi \big\rangle_\rho$ in the context of an expectation value with respect to $U(1)$ complexifier coherent states \cite{Bahr:Coherent,thiemann2006complexifier}. This observation is based on the fact that, as mentioned beforehand, there is no order-by-order expansion of the holonomies, which invalidates the strategy used for the solution of $\big\langle \hat{p}_\varphi \big\rangle_\rho$. Hereby we can use that at lowest order in the semiclassicality parameter, the operators reduce to their classical counterparts and quantum corrections occur in higher orders of the semiclassicality parameter only that shall be neglected here, enabling a straightforward solution of Equation (\ref{eq:eff_Q}). Direct integration then yields

\begin{align*}
    \big\langle \hat{Q}_\lambda \big\rangle_\rho(t) &= \sin\bigg( C_1 
    - \frac{1}{4} (t-t_0) E\bigg(\frac{3}{4}, \frac{G(t-t_0)^4}{24m_0^2}\bigg) + \frac{1}{2} \sqrt[4]{\frac{3m_0^2}{2G}} \Gamma\bigg(\frac{1}{4}\bigg) \bigg) \\[0.5em]
    &= \sin\bigg( C_1 
    - \frac{1}{4} \bigg( \frac{G}{24m_0^2} \bigg)^{-\frac{1}{4}} \; \Gamma\bigg(\frac{1}{4}, \frac{G(t-t_0)^4}{24m_0^2} \bigg) + \frac{1}{2} \sqrt[4]{\frac{3m_0^2}{2G}} \Gamma\bigg(\frac{1}{4}\bigg) \bigg),
\end{align*}
where $\lambda=1$ has been chosen. Moreover, $\big\langle \hat{Q}_\lambda \big\rangle_\rho(t_0) \defeq \sin\big( C_1\big)$ with $C_1 \in \mathbb{R}$ in the general case, $E(n,x)$ is the exponential integral function, closely related to the incomplete Gamma function $\Gamma(\cdot, \cdot)$ (with \textit{two} arguments, notably) and $\Gamma(\cdot)$ is the ordinary Gamma function, respectively. A visualization for an exemplary choice of parameters can be seen in Figure \ref{fig:plot_p} and Figure \ref{fig:plot_q} below, respectively.

\begin{figure}[h]
\centering
\includegraphics[scale=0.9]{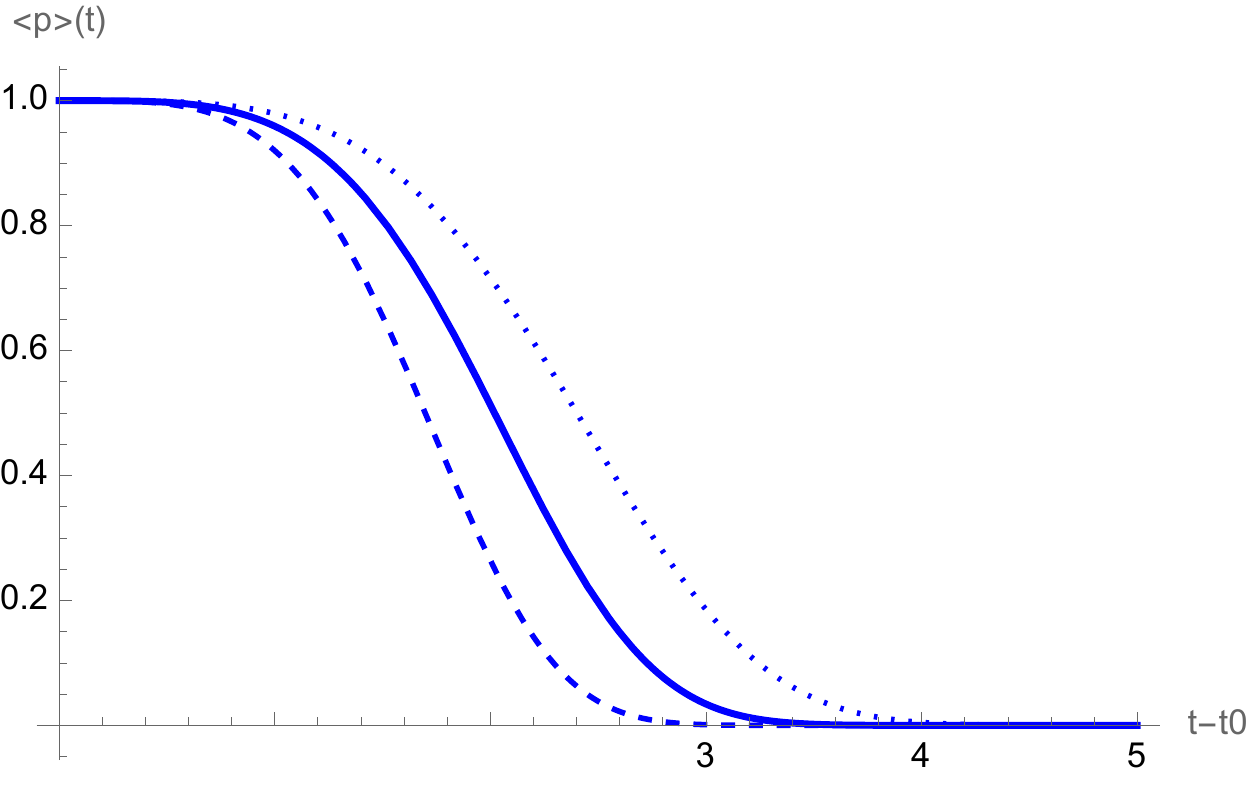}
\caption{Visual illustration of the behaviour of the solution $\big\langle \hat{p}_\varphi \big\rangle_\rho(t)$ for $\frac{G}{m_0^2} = \big\{ \frac{1}{2}, 1, 2 \big\}$ represented as the dotted, thick and dashed lines, respectively. The~parameter values are chosen as $\big\langle \hat{p}_\varphi \big\rangle_\rho(t_0)=1$ and $t_0 = 0$ for convenience. }
\label{fig:plot_p}
\end{figure}

\begin{figure}[H]
\centering
\includegraphics[scale=0.9]{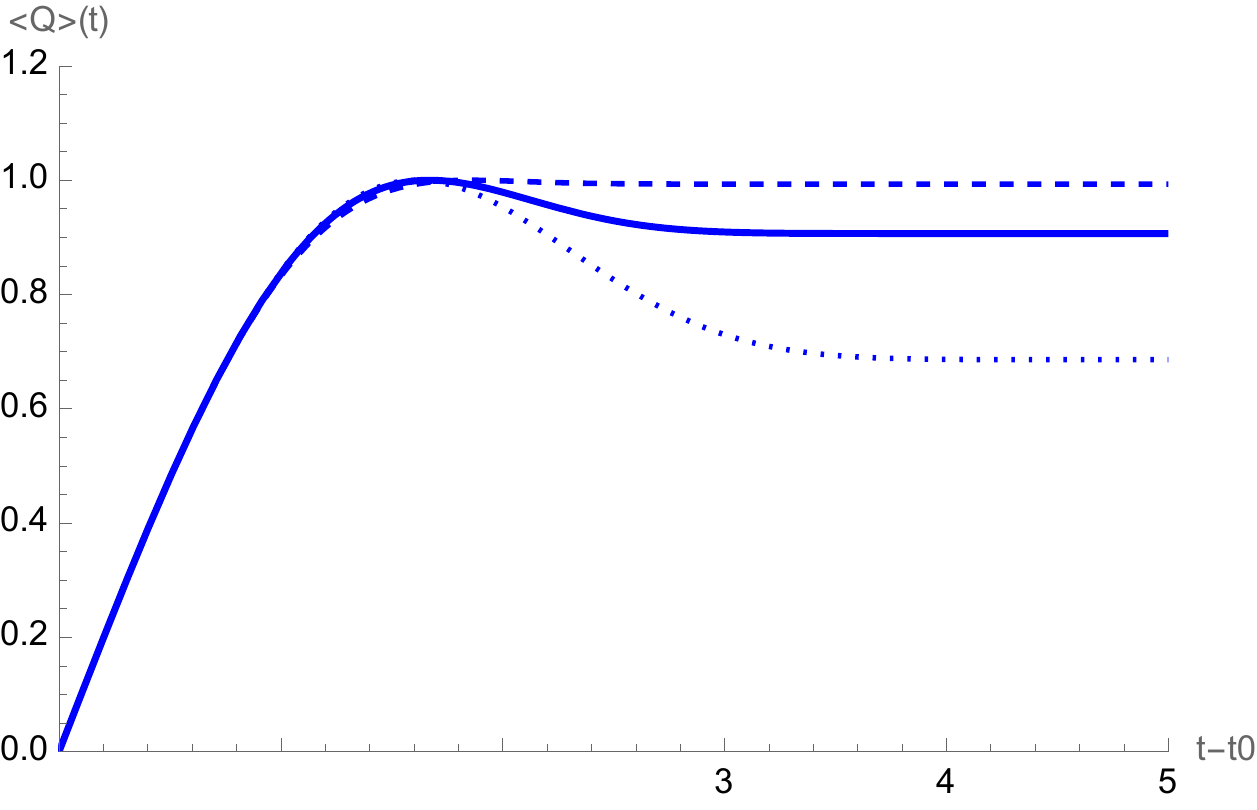}
\caption{Visual illustration of the behaviour of the solution $\big\langle \hat{Q}_\lambda \big\rangle_\rho(t)$  for $\lambda=1$ and the lowest semiclassicality order in $U(1)$ complexifier coherent states with $\frac{G}{m_0^2} = \big\{ \frac{1}{2}, 1, 2 \big\}$ represented as the dotted, thick and dashed lines, respectively. These are in direct correspondence to Figure \ref{fig:plot_p} above. The~parameter values are chosen as $\big\langle \hat{Q}_\lambda \big\rangle_\rho(t_0)=0$ and $t_0 = 0$ for convenience.}
\label{fig:plot_q}
\end{figure}

Corresponding to the solution for the momentum expectation value in (\ref{eq:solution_p}) and visualized in Figure \ref{fig:plot_p}, the lowest-order behaviour of $\big\langle \hat{Q}_\lambda \big\rangle_\rho(t)$ reflects the strong momentum damping in the sense that the motion of the polymer particle comes to a halt after a characteristic timescale defined by the relation of coupling constants $g_\alpha$ and masses $m_0$. We would like to stress, however, that the above relations between coupling parameter and particle mass are purely exemplary and should be considered a proof of concept. In order to obtain realistic predictions, we would need to make sure to avoid Poincaré recurrences by virtue of an introduction of a proper continuum limit \cite{Breuer:OpenSystems} regarding the environmental degrees of freedom. The~choice of a spectral density with appropriate physical properties to effectively describe the environment in the context of a polymerized model is still an open issue to the best knowledge of the authors.

\section{Conclusions} \label{sec:conclusion}
In this article we derived and investigated an open U(1) scattering model in the framework of polymerized quantum mechanics \cite{Ashtekar:Polymer,corichi2007polymer,corichi2007hamiltonian} which mimics collisional decoherence that can be regarded as one of the simplest decoherence models. One of our motivations to consider such a model is to obtain a better understanding of how decoherence effects look like in polymerized quantum mechanics as compared to their corresponding Schrödinger-like models. This is a first step towards the ability to analyze these effects also in more complicated LQC- and/or LQG-inspired models at a later stage. Next to its simplicity it is also an interesting model because using a Schr\"odinger representation, as has been done in \cite{hornberger2003collisional,hornberger2006master,busse2009pointer,hornberger2008monitoring}, the usual approach, to start with a given total Hamiltonian that involves the dynamics of the system and environment as well as its interaction, does not work in a straightforward manner here because one commonly encounters singularities in the derivation of the master equation (more precisely the correlation functions) for such a model \cite{hornberger2003collisional}. The~main difference in the setup considered here compared to previous work on collisional decoherence in \cite{hornberger2003collisional,hornberger2006master,busse2009pointer,hornberger2008monitoring} is that in polymerized quantum mechanics, depending on the chosen specific representation either only the  position or momentum operator exist on the polymer Hilbert space. For the remaining elementary variable, only the corresponding exponentiated version can be implemented as an operator but not the corresponding generator. A similar situation is for instance also considered in \cite{Kastrup:AnglePair,Kastrup:Optical} for the quantization of the phase space in applications of quantum optics. In the notation of \cite{corichi2007polymer}, we have chosen the $A$-type representation where the momentum operator exists. As a consequence, operators involving the position need to be expressed in terms of the associated Weyl elements. The~latter allows to base the model on a chosen total Hamiltonian that involves no potential but involves the polymerized version of typical interaction Hamiltonians leading to decoherence and dissipation. Furthermore, our choice was guided by the results for Lindblad operators found in previous works on the subject \cite{hornberger2003collisional,hornberger2006master,busse2009pointer,hornberger2008monitoring}.
\\

Our results deviate from and extend the results present in the literature in several aspects. Firstly, we utilized the projection operator formalism in the form of the TCL approach in order to formulate a master equation, whereas \cite{hornberger2003collisional,hornberger2006master,hornberger2008monitoring} implemented an S-matrix-like strategy for the scattering of a massive Brownian particle in an ideal gas environment. Secondly, due to the different representation of the canonical variables, our Hilbert space looks different in the sense that our basis functions do not belong to the standard Hilbert space $L^2(\mathbb{R}, \d q)$, where they would be non-normalizable. The~Gibbs state in the form of the environmental density operator consequently has a different form than the one used previously, which is one aspect of the avoidance of the afore-mentioned singularities. These can however also be circumvented in Schrödinger quantum mechanics by the choice of an overcomplete decomposition of the density operator as discussed in~\cite{hornberger2003collisional}, albeit not in the context of a TCL master equation. 
\\

In order to analyze the physical properties of the derived master equation we considered a Taylor series in the polymerization parameter, where we only apply  the expansion to functions, not operators involved in the final master equation. Our results show that the lowest-order solution of the effective differential equation for the momentum operator expectation value in the case of equal system and environmental mass reveals a strong momentum damping based on the relation between coupling strength and the value of the particle masses. It is noteworthy that, with a non-vanishing coupling and finite particle mass, we cannot reproduce $\langle \hat{p}_\varphi \rangle_\rho = \textrm{const.}$ as is the case in \cite{hornberger2006master} for a symmetrical momentum transfer function. This holds even in zeroth order in the polymerization parameter $\lambda$, where one intuitively might expect the result from Schrödinger quantum mechanics as a limiting case. Taking the continuum limit is however not as trivial as sending the lattice spacing to zero \cite{corichi2007polymer}, the Hilbert space used in the model presented here has a manifestly different structure than just $L^2(\mathbb{R}, \d q)$. In this sense the results for $\langle \hat{p}_\varphi \rangle_\rho$ and $\langle \hat{Q}_\lambda \rangle_\rho$ are a footprint of the structure of the polymer Hilbert space and its associated superselection sectors. Solving the (then coupled) differential equations between these two expectation values up to higher orders in $\lambda$ and comparing these results to both the existing solutions for Schrödinger quantum mechanics and the lowest-order approximation is the goal of a future analysis of this class of models. For the purpose of a more direct comparison, we would need to analyze the results of an analogous interaction Hamiltonian in terms of holonomies on the Schrödinger Hilbert space, highlighting the structural differences between Hilbert spaces. Up to the correlation functions the results would be identical due to the observable algebra, at the level of the correlation functions, the question arises whether certain convex decompositions of the density operator are preferred in the sense of the avoidance of singularities as for example pointed out in \cite{hornberger2003collisional}. This will be analyzed in detail in the course of a future project.
\\

At the level of the master equation discussed here, as well as the effective equations for the expectation values of the momentum operator and the polymerized position are concerned, it is clear that the second Markov approximation cannot be performed without any further assumptions as well. These assumptions can for example encompass the choice of a spectral density as a replacement for the sum over coupling constants and environmental particle labels and corresponds in a sense to the environmental thermodynamic limit. This replacement can be intuitively understood as regarding the particle number in the environment as sufficiently large such that it can be approximated by a continuum of modes (or as in our case, masses) rather than a discrete sum \cite{Breuer:OpenSystems,Schlosshauer:DecoherenceBook}. In the course of this approximation it is convenient to replace the coupling by a frequency- or mass-dependent function with a defined cutoff, such as a Lorentzian or a (sub-)Ohmic spectral density, see for example \cite{Breuer:OpenSystems,Schlosshauer:DecoherenceBook,clos2012quantification,lambert2019modelling}. Consequently, all quantities including the environmental particle mass are part of the integrand. With this method, depending on the explicit form of the spectral density and the correlation functions, the second Markov approximation might be employed, which leads to time-independent operators in the Born--Markov master equation. In the model considered here  however, we did not replace the couplings by a spectral density and hence kept the explicit time-dependence in the final equations. Since this time-dependence is crucial for the final result presented in this work, it is clear that an application of too many approximations alters the physical predictions of a given model, such as for example already encountered in \cite{feller2017surface} for a first decoherence model regarding surface states in LQG. 
\\

A valid question to ask is whether the model can be generalized in a way involving a combination of Hamiltonians and associated couplings such that the Lindblad operators found in Schrödinger collisional decoherence are directly obtained in the derivation of a Born--Markov master equation. In the vicinity of the model we discussed in this work, the answer is in the negative. The~coupling scheme we used here corresponds to the polymerized version of an interaction of the form $q \otimes q$ which is, among the four possible choices, the most complicated one as far as the derivation is considered. There are three remaining different coupling schemes that could have been used in the interaction Hamiltonian. A $q$-polymerized coupling in the interaction Hamiltonian of the form $q \otimes p$ would lead to time-independent correlation functions. As a consequence, this simply leads to a time-integral of the interaction picture of the system's interaction Hamiltonian in the master equation. It is at this point not obvious how this can be performed at the operator level in a meaningful way. The~second option would be an interaction term of the form $p \otimes q$, which yields identical environmental correlation functions but time-independent system operators in the, now trivial, interaction picture. The~resulting temporal integral of the trigonometric functions in the master equation can be in principle performed, the second Markov approximation is however not applicable due to the ill-definedness of the integral in that case. Thirdly, an interaction of the form $p \otimes p$ contains no polymerized quantities and has an entirely trivial interaction picture. Consequently, the integral in the Born--Markov master equation gives rise to a linear time-dependence as for example in the case of \cite{feller2017surface}. The~limit $t_0 \to -\infty$ is not applicable as well, hence the second part of the Markov approximation relying on the peakedness of the environmental correlation functions or the choice of a suitable spectral density fails. 
\\

The~presented results also provide a starting point towards the investigation of the behaviour of (complexifier) coherent states \cite{thiemann2006complexifier,Bahr:Coherent} and the decoherence-free Hilbert space sectors of so-called pointer states \cite{anglin1996decoherence,zurek2003decoherence,dalvit2005predictability}. The~dissipator obtained in the final master equation of the open U(1) scattering model in (\ref{eq:U1_master}) in Section \ref{sec:model_application} can in principle be directly applied to existing $U(1)$ coherent states, revealing the compatibility of these states with the notion of classicality possibly brought about by the master equation derived here. It was shown in \cite{busse2009pointer} that the pointer states for collisional decoherence presented in their work consists of solitonic states. In this context, an interesting albeit challenging task would be to find a suitable generalization of soliton states that is applicable to the polymerized model in this article and gives rise to a decoherence-free subspace of a superselected polymer Hilbert space. Analogously, the presented computations or rather a variant thereof can be performed in the $B$-type representation (see \cite{corichi2007polymer} for the respective nomenclature). The~crucial difference is that the background dynamics would be generated by a combination of holonomies where the interaction Hamiltonian would be given by the proper position operators (although with a discrete spectrum due to momentum polymerization). This brings along the difficulty of defining an initial state (most prominently a Gibbs state) in terms of an eigenbasis decomposition of a combination of holonomies, whereas the interaction picture transformation turns out to be even easier than in the case presented~above. 
\\

A similarly structured model with bilinear interaction Hamiltonian can in the future be implemented for the polymer quantum harmonic oscillator, which would then be the polymerized counterpart of QBM or the Caldeira--Leggett model \cite{Caldeira:QBM} in the case of a $q \otimes q$-type interaction. Based on our considerations, it is immediately clear that this would be more complex in two aspects: On the one hand this involves the transition to the interaction picture, since the background Hamiltonians for the system and environment include both polymerized and non-polymerized operators. This complicates the solution of the Heisenberg equation or the closed summation of the iterated commutators, respectively. On the other hand, since the eigenfunctions of the closed polymer quantum harmonic oscillator are the Mathieu functions \cite{Ashtekar:Polymer}, the explicit computation of the environmental correlation functions might require perturbative methods. The~process of discussing gradually more and more complex models will be useful for the long-term goal to consider open quantum models also in the context of symmetry-reduced, loop-quantized models such as open LQC as well as in the context of loop quantized gravitationally-induced decoherence models as extensions of the model presented in \cite{Fahn:2022zql}.

\acknowledgments
M.K thanks the Heinrich-Böll foundation and the Friedrich-Alexander Universität for financial support.

\appendix
\section{Quantization of a Bead on a Circle} \label{sec:bead_circle}
The~configuration manifold in standard quantum mechanics usually is some Cartesian product of the real line, that is $\mathbb{R}, \mathbb{R}^2$ and so on. This fact of course entails the unboundedness of position and momentum operators, respectively, whereas the representations of the (bounded) Weyl elements are unique up to unitary equivalence. In the context of quantum gravity \cite{Thiemann:MCQGR} or more specifically quantum cosmology \cite{Ashtekar:LQCReport,Agullo:LQCReview}, the quantization procedures deviate drastically from the usual approach in the sense that the relevant quantities are not simply the configuration degrees of freedom and their associated momenta but generally more complicated objects such as holonomies of a given gauge group. In the course of this formalism, von Neumann's uniqueness result for weakly continuous one-parameter unitary groups is violated, which naturally comes with the major caveat of finding an appropriate representation. This is, albeit in a less complicated manner, also the case for a particle with a configuration manifold topologically equivalent to a circle $S^1$. In this case, the angle variable $\varphi$ is not a suitable variable for quantization as it fails to exhibit a key feature of the theory's phase space, periodicity. A quantization of the associated Weyl element (or holonomy, respectively) is much more approriate and leads exactly to a representation space equal to the superselection Hilbert spaces we encountered in the main part of this article. The~prime example of a particle with a configuration manifold equal to $S^1$ (or $U(1)$ if a group action is implied) is a bead with mass $m_0$ moving frictionlessly on a circular wire with some fixed radius $r_0$. The~associated Lagrangian and Hamiltonian, respectively, are easily found:

\begin{equation}
    L = \frac{1}{2} m_0 \Dot{\varphi}^2 r_0^2, \quad
    p_\varphi = m_0 r_0^2 \Dot{\varphi}, \quad 
    H = \frac{p_\varphi^2}{2 m r_0^2}, \quad
    \Dot{\varphi} = \{\varphi, H\}_{(\varphi, p_\varphi)} = \frac{p_\phi}{m_0 r_0^2}=\omega = \textrm{const.}
\end{equation}

It is immediately clear that the phase space of this system has the topology of $S^1 \times \mathbb{R}$ (as the cotangent bundle of the circle) with the configuration variable $\varphi \in [0, 2\pi)$ and $p_\varphi \in \mathbb{R}$ being a constant of motion that purely depends on the initial conditions. Consequently, it is impractical to implement the quantization procedure with the canonical pair $(\varphi, p_\varphi)$, we rather need variables that reflect the structure of the phase space. This is achieved by the consideration of complex exponentials of the angle coordinate. A similar procedure is implemented in \cite{Kastrup:AnglePair,Kastrup:Optical} in the context of the optical phase space with an angle variable and its canonical (angular) momentum, although with real variables instead of complex ones as in our case. We will loosely follow the course of these works here. That is, our new canonical coordinates are given by the triple $\big( e^{i \varphi}, e^{-i \varphi}, p_\varphi \big)$ with a Poisson structure according to:

\begin{equation*}
    \{ e^{ \pm i \varphi}, p_\varphi \} = \bigg( \frac{\d}{\d \varphi} e^{ \pm i \varphi} \bigg) \frac{\d p_\varphi}{\d p_\varphi} - \frac{\d p_\varphi}{\d \varphi} \bigg( \frac{\d}{\d p_\varphi} e^{ \pm i \varphi} \bigg) = \pm i e^{ \pm i \varphi}, \quad \{ e^{ \pm i \varphi}, e^{ \mp i \varphi} \} = 0. 
\end{equation*}

Intuitively, this Poisson algebra contains the generators of (complex) rotations and translations, hence it corresponds to the Lie algebra of the Euclidean group $E(2)$ of the two-dimensional plane, which in itself consists of subgroups for rotations and translations, respectively. This is especially easy to see in a rotated coordinate system with $\sin(\varphi), \cos(\varphi), p_\varphi$ as the generators. Upon quantization, this structure is expressed in terms of commutators within a $C^*$-algebra of the above variables, where for brevity we set $U \defeq e^{i \varphi}$ and $U^+ \defeq e^{-i \varphi}$ previous to quantization:

\begin{equation}
    [\hat{U}, \hat{p}_\varphi] 
    = i \widehat{ \{U, p_\varphi \} } = -\hat{U}, \quad
    [\hat{U}^\dagger, \hat{p}_\varphi] 
    = i \widehat{ \{U^+, p_\varphi \} } = \hat{U}^\dagger, \quad [\hat{U}, \hat{U}^\dagger] = 0.
    \label{eq:polymer_comm}
\end{equation}

As a next step we need to provide a representation space with a suitable inner product. It turns out that all irreducible unitary representations of $E(2)$ and its coverings \cite{Kastrup:AnglePair} can be implemented on $L^2\big(S^1, \tfrac{\d \varphi}{2 \pi}\big)$ with the standard inner product, labelled by two parameters $r > 0$ and $\delta \in [0,1)$. In our case $\delta$ is the important of the two as it essentially determines if the representation corresponds to the group itself ($\delta = 0$), a $q$-fold covering group ($\delta \in \mathbb{Q}$) or the universal covering ($\delta \in \mathbb{R} \setminus \mathbb{Q}$) of $E(2)$. The~physical consequences of the numerical value of $\delta$ will be discussed in the course of this text. The~representation space with the chosen inner product has the convenient orthonormal basis $e_n(\varphi) = e^{i n \varphi}$ with $n \in \mathbb{Z}$. This leads to the self-adjoint generator of rotations $\hat{p}^{(\delta)}_\varphi = -i \partial_\varphi + \delta$ in the 'position' representation, whereas $\hat{U}, \hat{U}^\dagger$ are not self-adjoint but unitary and act multiplicatively. The~important distinction between this case and a proper loop representation is that the operators $\hat{U}, \hat{U}^\dagger$ are \textit{weakly} continuous, in otherwords there is an infinitesimal, self-adjoint generator in the form of an \textit{angle operator} $\hat{\varphi}$, we hence still operate within the Stone-von Neumann theorem on uniqueness up to unitary equivalence akin to Schrödinger quantum mechanics. Note however that $\hat{\varphi}$ does not correspond to a function on phase space due to the lack of periodicity. Conveniently, the basis state $e_n(\varphi)$ are eigenstates of the angular momentum operator with eigenvalues $(n + \delta)$. Of course, different values of $\delta$ lead to different spectra of $\hat{p}^{(\delta)}_\varphi$, the individual representations are hence not unitarily equivalent. It is worth noting that this particular representation of the momentum has similarities to a covariant derivative and indeed can be interpreted accordingly in the context of non-trivial phase shifts associated with the Aharonov-Bohm effect. Although different values of $\delta$ lead to unitarily inequivalent representations, it turns out that we can shift this parameter to the Hilbert space basis instead of having to deal with it in terms of the operator.

\begin{equation*}
   L^2\big(S^1, \tfrac{\d \varphi}{2 \pi}\big) \to 
   L^2\big(S^1, \tfrac{\d \varphi}{2 \pi}, \delta \big), \quad
   e_n(\varphi) \mapsto e_n^{(\delta)}(\varphi) \defeq e^{i (n+\delta) \varphi} \textrm{ with } n \in \mathbb{Z}, \quad \hat{p}_\varphi = -i \partial_\varphi.
\end{equation*}

Naturally, the basis and all functions expanded in that basis accrue a phase factor with every full rotation, that is $e_n^{(\delta)}(\varphi + 2\pi) = e^{i 2\pi  \delta} e_n^{(\delta)}(\varphi)$. Note that the angular momentum operator in this particular quantization is not bounded from below, contrary to the case of a cone-like (instead of a cylindrically shaped) phase space geometry associated with a similarly quantized harmonic oscillator Hamiltonian \cite{Kastrup:Optical}.

\section{Interaction Picture: An Alternative Route} \label{sec:appendixB}
A different route to the constituents of the interaction picture interaction Hamiltonian (\ref{eq:int_picture_U}) and (\ref{eq:int_picture_Udagger}) is given by the iterated commutator bracket or the adjoint action of the free particle (background) Hamiltonian respectively. This method of transforming into the interaction picture might be easier or harder than comparable approaches, depending on the details of the model at hand. In our case,  we would like to evaluate the following~expression:

\begin{align*}
    \hat{U}^{(\alpha)}_\lambda (\xi) = \sum_{n=0}^\infty \frac{(i \xi)^n}{n!} \frac{1}{(2m_\alpha)^n} \Big[ \big( \hat{p}_\varphi^{(\alpha)} \big)^2, \hat{U}^{(\alpha)}_\lambda \Big]_{(n)}, \quad \big[ \hat{A},\hat{B} \big]_{(n)} \defeq \big[ \hat{A}, \big[ \hat{A},\hat{B}\big]_{(n-1)}\big], \quad \big[ \hat{A},\hat{B} \big]_{(0)} \defeq \hat{B}.
\end{align*}

In order to solve this infinite sum, it is helpful to have a closer look at the first few orders of the commutator. This way one can oftentimes make an educated guess regarding the higher-order terms, which can in turn then be proven inductively. For pragmatic purposes we would like to have a normal-ordered (or anti-normal-ordered) expression since this considerably simplifies consequent computations. For the sake of simplicity and due to the fact that operators labeled with $\alpha$ only act on the respective subspaces, we can omit this label for increased readability. Keeping in mind that the commutator amounts to $\big[ \hat{U}_\lambda, \hat{p}_\varphi \big] = - \lambda \hat{U}_\lambda$, we conclude:

\begin{align*}
    \Big[ \big( \hat{p}_\varphi \big)^2, \hat{U}_\lambda \Big]_{(1)} &= \hat{p}_\varphi \big[ \hat{p}_\varphi, \hat{U}_\lambda \big] - \big[ \hat{U}_\lambda, \hat{p}_\varphi \big] \hat{p}_\varphi = \lambda \big( \hat{p}_\varphi \hat{U}_\lambda + \hat{U}_\lambda \hat{p}_\varphi \big) = \lambda^2 \hat{U}_\lambda + 2 \lambda \hat{U}_\lambda \hat{p}_\varphi, \\[0.5em]
    \Big[ \big( \hat{p}_\varphi \big)^2, \hat{U}_\lambda \Big]_{(2)} &= \lambda \Big( \big[ \hat{p}_\varphi^2, \hat{U}_\lambda \big] \hat{p}_\varphi + \hat{p}_\varphi \big[ \hat{p}_\varphi^2, \hat{U}_\lambda \big] \Big)
    = \lambda^2 \Big( \hat{U}_\lambda \hat{p}_\varphi^2 + 2 \hat{p}_\varphi \hat{U}_\lambda \hat{p}_\varphi + \hat{p}_\varphi^2 \hat{U}_\lambda \Big) \\[0.1em]
    &= \lambda^2 \Big( \hat{U}_\lambda \hat{p}_\varphi^2 + 2 \big( \hat{U}_\lambda \hat{p}_\varphi - \big[\hat{U}_\lambda, \hat{p}_\varphi \big] \big) \hat{p}_\varphi + \hat{U}_\lambda \hat{p}_\varphi^2 - \big[ \hat{U}_\lambda, \hat{p}_\varphi^2 \big] \Big) \\[0.3em]
    &= \lambda^4 \hat{U}_\lambda + 4 \lambda^3 \hat{U}_\lambda \hat{p}_\varphi + 4 \lambda^2 \hat{U}_\lambda \hat{p}_\varphi^2, \\
    & \; \; \vdots \\
    \Big[ \big( \hat{p}_\varphi \big)^2, \hat{U}_\lambda \Big]_{(n)} &= \lambda^n \sum_{k=0}^n \binom{n}{k} \hat{p}_\varphi^k \, \hat{U}_\lambda \, \hat{p}_\varphi^{n-k}
    = \sum_{k=0}^n \sum_{l=0}^k \binom{n}{k} \binom{k}{l} \lambda^{n+k-l} \, \hat{U}_\lambda \, \hat{p}_\varphi^{n-k+l}.
\end{align*}

This can straightforwardly be proven by induction in two steps.

\begin{align*}
   \Big[ \big( \hat{p}_\varphi \big)^2, \hat{U}_\lambda \Big]_{(n+1)} &= \Big[ \big(\hat{p}_\varphi \big)^2, \Big[ \big( \hat{p}_\varphi \big)^2, \hat{U}_\lambda \Big]_{(n)} \Big]
   = \lambda^n \sum_{k=0}^n \binom{n}{k} \hat{p}_\varphi^k \, \Big[ \big( \hat{p}_\varphi \big)^2, \hat{U}_\lambda \Big] \, \hat{p}_\varphi^{n-k} \\[0.3em]
   &= \lambda^{n+1} \bigg( \sum_{k=0}^n \binom{n}{k} \hat{p}_\varphi^k \, \hat{U}_\lambda \, \hat{p}_\varphi^{n-k+1} + \sum_{k=0}^n \binom{n}{k} \hat{p}_\varphi^{k+1} \, \hat{U}_\lambda \, \hat{p}_\varphi^{n-k} \bigg) \\[0.3em]
   &= \lambda^{n+1} \bigg( \sum_{k=0}^n \binom{n}{k} \hat{p}_\varphi^k \, \hat{U}_\lambda \, \hat{p}_\varphi^{n-k+1} + \sum_{k=1}^{n+1} \binom{n}{k-1} \hat{p}_\varphi^{k} \, \hat{U}_\lambda \, \hat{p}_\varphi^{n-k+1} \bigg) \\[0.3em]
   &= \lambda^{n+1} \bigg( \sum_{k=1}^n \bigg[ \binom{n}{k} + \binom{n}{k-1} \bigg] \hat{p}_\varphi^{k} \, \hat{U}_\lambda \, \hat{p}_\varphi^{n-k+1} + \hat{U}_\lambda \hat{p}_\varphi^{n+1}+ \hat{p}_\varphi^{n+1} \hat{U}_\lambda \bigg) \\[0.3em]
   &= \lambda^{n+1} \bigg( \sum_{k=1}^n \binom{n+1}{k} \hat{p}_\varphi^{k} \, \hat{U}_\lambda \, \hat{p}_\varphi^{n-k+1} + \hat{U}_\lambda \hat{p}_\varphi^{n+1}+ \hat{p}_\varphi^{n+1} \hat{U}_\lambda \bigg) \\[0.3em]
   &= \lambda^{n+1} \sum_{k=0}^{n+1} \binom{n+1}{k} \hat{p}_\varphi^{k} \, \hat{U}_\lambda \, \hat{p}_\varphi^{n-k+1}.
\end{align*}

Normal-ordering the holonomies then consequently yields:

\begin{align*}
    \hat{p}_\varphi^{k+1} \hat{U}_\lambda &= \hat{p}_\varphi \hat{p}_\varphi^{k} \, \hat{U}_\lambda = \sum_{l=0}^k \binom{k}{l} \lambda^{k-l} \hat{p}_\varphi\, \hat{U}_\lambda \, \hat{p}_\varphi^l 
    = \sum_{l=0}^k \binom{k}{l} \lambda^{k-l} \bigg( \Big[\hat{p}_\varphi, \hat{U}_\lambda \Big] + \hat{U}_\lambda \, \hat{p}_\varphi \bigg) \hat{p}_\varphi^l \\[0.3em]
    &= \sum_{l=0}^k \binom{k}{l} \lambda^{k-l} \bigg( \lambda \hat{U}_\lambda \, \hat{p}_\varphi^l + \hat{U}_\lambda \, \hat{p}_\varphi^{l+1} \bigg)
    = \sum_{l=0}^k \binom{k}{l} \lambda^{k-l} \hat{U}_\lambda \, \hat{p}_\varphi^l + \sum_{l=1}^{k+1} \binom{k}{l-1} \lambda^{k-l+1} \hat{U}_\lambda \, \hat{p}_\varphi^l \\[0.3em]
    &= \sum_{l=1}^{k+1} \bigg[ \binom{k}{l} + \binom{k}{l-1} \bigg] \lambda^{k-l+1} \hat{U}_\lambda \, \hat{p}_\varphi^l + \lambda^{k+1} \hat{U}_\lambda + \hat{U}_\lambda \hat{p}_\varphi^{k+1}
    = \sum_{l=0}^{k+1} \binom{k+1}{l} \lambda^{k-l+1} \hat{U}_\lambda \, \hat{p}_\varphi^l.
\end{align*}

A fully analogous calculation for the interaction picture of the adjoint $\hat{U}^\dagger_\lambda$ yields to a result that only differs by a minus sign. This can be easily seen if we consider the iterated commutator for this case and recall that $\big[ \hat{U}^\dagger_\lambda, \hat{p}_\varphi \big] = \lambda \hat{U}^\dagger_\lambda$:

\begin{align*}
    \Big[ \big( \hat{p}_\varphi \big)^2, \hat{U}^\dagger_\lambda \Big]_{(1)} &= \hat{p}_\varphi \big[ \hat{p}_\varphi, \hat{U}^\dagger_\lambda \big] - \big[ \hat{U}^\dagger_\lambda, \hat{p}_\varphi \big] \hat{p}_\varphi = - \lambda \big( \hat{p}_\varphi \hat{U}^\dagger_\lambda + \hat{U}^\dagger_\lambda \hat{p}_\varphi \big) = \lambda^2 \hat{U}^\dagger_\lambda - 2 \lambda \hat{U}^\dagger_\lambda \hat{p}_\varphi, \\[0.5em]
    \Big[ \big( \hat{p}_\varphi \big)^2, \hat{U}^\dagger_\lambda \Big]_{(2)} &= - \lambda \Big( \big[ \hat{p}_\varphi^2, \hat{U}^\dagger_\lambda \big] \hat{p}_\varphi + \hat{p}_\varphi \big[ \hat{p}_\varphi^2, \hat{U}^\dagger_\lambda \big] \Big)
    = - \lambda \Big( - \lambda \hat{U}^\dagger_\lambda \hat{p}^2_\varphi - 2 \lambda \hat{p}_\varphi \hat{U}^\dagger_\lambda \hat{p}_\varphi - \lambda \hat{p}^2_\varphi \hat{U}^\dagger_\lambda \Big) \\[0.4em]
    &= \lambda^2 \Big( \hat{U}^\dagger_\lambda \hat{p}^2_\varphi + 2  \hat{p}_\varphi \hat{U}^\dagger_\lambda \hat{p}_\varphi + \hat{p}^2_\varphi \hat{U}^\dagger_\lambda \Big)
    = \lambda^4 \hat{U}^\dagger_\lambda - 4 \lambda^3 \hat{U}^\dagger_\lambda \hat{p}_\varphi + 4 \lambda^2 \hat{U}^\dagger_\lambda \hat{p}^2_\varphi, \\
    & \; \; \vdots \\
    \Big[ \big( \hat{p}_\varphi \big)^2, \hat{U}^\dagger_\lambda \Big]_{(n)} &= \big( -\lambda \big)^n \sum_{k=0}^n \binom{n}{k} \hat{p}_\varphi^k \, \hat{U}_\lambda \, \hat{p}_\varphi^{n-k}
    = \sum_{k=0}^n \sum_{l=0}^k \binom{n}{k} \binom{k}{l} \big( -\lambda \big)^{n+k-l} \, \hat{U}_\lambda \, \hat{p}_\varphi^{n-k+l},
\end{align*}

where we observed that every instance of $\big[ \hat{p}_\varphi, \hat{U}^\dagger_\lambda \big]$ yields a factor of $- \lambda$ and eliminates a power of $\hat{p}_\varphi$ in the process. Apart from the alternating sign, the inductive proof is absolutely identical. Upon closer inspection we realize that these are the sum representations for an exponential. Since $\hat{U}_\lambda$ does not carry a summation index it can be moved to the very left of the expression, letting us evaluate the remaining sums explicitly. A similar procedure leads to the analogous result for $\hat{U}_\lambda(\xi)^\dagger$ but with a flipped sign in front of $\lambda$ due to the structure constants of the operator algebra. The~closed-form expressions are finally given by:

\begin{align*} \hspace{-0.3cm}
    \hat{U}_\lambda(\xi) &= \sum_{n=0}^\infty \frac{1}{n!} \frac{(i \xi)^n}{(2m_\alpha)^n} \sum_{k=0}^n \sum_{l=0}^k \binom{n}{k} \binom{k}{l} \lambda^{n+k-l} \, \hat{U}_\lambda \, \hat{p}_\varphi^{n-k+l} = \hat{U}_\lambda \exp{\frac{i \xi}{2m} \bigg( \big( \hat{p}_\varphi + \lambda \mathds{1} \big)^2 -  \hat{p}_\varphi^2 \bigg)}, \\[0.3em]
     \hat{U}_\lambda^\dagger(\xi) &= \sum_{n=0}^\infty \frac{1}{n!} \frac{(i \xi)^n}{(2m_\alpha)^n} \sum_{k=0}^n \sum_{l=0}^k \binom{n}{k} \binom{k}{l} \big( -\lambda \big)^{n+k-l} \, \hat{U}_\lambda^\dagger \, \hat{p}_\varphi^{n-k+l} = \hat{U}_\lambda^\dagger \exp{\frac{i \xi}{2m} \bigg( \big( \hat{p}_\varphi - \lambda \mathds{1} \big)^2 -  \hat{p}_\varphi^2 \bigg)}.
\end{align*}

\newpage
\section*{References}
\bibliographystyle{unsrt}
\bibliography{DecoherenceBib}

\end{document}